\documentclass[twocolumn,tighten]{aastex631}
\usepackage{CJK}
\usepackage{amsmath}
\usepackage{multirow}
\usepackage{listings}
\usepackage{xcolor}

\lstdefinelanguage{Pseudo}{
  morekeywords=[1]{repeat,for,until,fix,update,lower,return,each,in,initialize,is,if,and,converge},
  keywordstyle=[1]\color{blue}\bfseries,
  morekeywords=[2]{grid_search_cv,early_stopping,median_across_folds},
  keywordstyle=[2]\color{orange}\bfseries,
  sensitive=false,
}

\newcommand{\magauto}{\texttt{MAG\_AUTO}}
\newcommand{\mumax}{\texttt{MU\_MAX}}
\newcommand{\xgboost}{\texttt{XGBoost}}
\newcommand{\dr}[1]{DR{#1}}

\defcitealias{Tachibana_2018}{TM18}

\graphicspath{{./}{Figures/}}

\begin{document}
\begin{CJK*}{UTF8}{gbsn}
\title{A Morphological Model to Separate Resolved--unresolved Sources in the DESI Legacy Surveys: Application in the LS4 Alert Stream}

\author[0000-0002-7866-4531]{Chang~Liu (刘畅)}
\affil{Department of Physics and Astronomy, Northwestern University, 2145 Sheridan Rd, Evanston, IL 60208, USA}
\affil{Center for Interdisciplinary Exploration and Research in Astrophysics (CIERA), Northwestern University, 1800 Sherman Ave, Evanston, IL 60201, USA}

\author[0000-0001-9515-478X]{Adam~A.~Miller}
\affil{Department of Physics and Astronomy, Northwestern University, 2145 Sheridan Rd, Evanston, IL 60208, USA}
\affil{Center for Interdisciplinary Exploration and Research in Astrophysics (CIERA), Northwestern University, 1800 Sherman Ave, Evanston, IL 60201, USA}
\affil{NSF-Simons AI Institute for the Sky (SkAI), 172 E. Chestnut St., Chicago, IL 60611, USA}

\author[0000-0002-7777-216X]{Joshua~S.~Bloom} 
\affil{Department of Astronomy, University of California, Berkeley, 501 Campbell Hall, Berkeley, CA 94720, USA}
\affil{Lawrence Berkeley National Laboratory, 1 Cyclotron Road, MS 50B-4206, Berkeley, CA 94720, USA}

\author[0000-0002-3803-1641]{Robert~A.~Knop} \affiliation{Lawrence Berkeley National Laboratory, 1 Cyclotron Road, MS 50B-4206, Berkeley, CA 94720, USA}

\author[0000-0002-3389-0586]{Peter~E.~Nugent}
\affil{Lawrence Berkeley National Laboratory, 1 Cyclotron Road, MS 50B-4206, Berkeley, CA 94720, USA}
\affil{Department of Astronomy, University of California, Berkeley, 501 Campbell Hall, Berkeley, CA 94720, USA}



\begin{abstract}
Separating resolved and unresolved sources in large imaging surveys is a fundamental step to enable downstream science, such as searching for extragalactic transients in wide-field time-domain surveys. Here we present our method to effectively separate point sources from the resolved, extended sources in the Dark Energy Spectroscopic Instrument (DESI) Legacy Surveys (LS). We develop a supervised machine-learning model based on the Gradient Boosting algorithm \texttt{XGBoost}. The features input to the model are purely morphological and are derived from the tabulated LS data products. We train the model using $\sim$$2\times10^5$ LS sources in the COSMOS field with HST morphological labels and evaluate the model performance on LS sources with spectroscopic classification from the DESI Data Release 1 ($\sim$$2\times10^7$ objects) and the Sloan Digital Sky Survey Data Release 17 ($\sim$$3\times10^6$ objects), as well as on $\sim$$2\times10^8$ Gaia stars. A significant fraction of LS sources are not observed in every LS filter, and we therefore build a ``Hybrid'' model as a linear combination of two \texttt{XGBoost} models, each containing features combining aperture flux measurements from the ``blue'' ($gr$) and ``red'' ($iz$) filters. The Hybrid model shows a reasonable balance between sensitivity and robustness, and achieves higher accuracy and flexibility compared to the LS morphological typing. With the Hybrid model, we provide classification scores for $\sim$$3\times10^9$ LS sources, making this the largest ever machine-learning catalog separating resolved and unresolved sources. The catalog has been incorporated into the real-time pipeline of the La Silla Schmidt Southern Survey (LS4), enabling the identification of extragalactic transients within the LS4 alert stream.
\end{abstract}

\keywords{Catalogs(205) --- Classification(1907) --- Surveys(1671) --- Astrostatistics(1882)}


\section{Introduction} \label{sec:intro}

Many wide-field imaging surveys have now cataloged $>$$10^9$ objects, including the Panoramic Survey Telescope and Rapid Response System \citep[Pan-STARRS;][]{PanSTARRS_2016}, the Dark Energy Spectroscopic Instrument (DESI) Legacy Imaging Surveys \citep[LS;][]{LS_2019}, and Gaia \citep{Gaia_2016,GaiaDR3_2023}. Object classification, e.g., separating foreground stars from distant galaxies, is a critical step for analyzing the sources observed by these surveys. Spectroscopic classifications are only available for $<$1\% of the imaged sources \citep{DESI_DR1_2025}. A classification framework based solely on imaging data is the only feasible way to separate the billions of resolved sources (e.g., galaxies and nebulae) from unresolved, point-like sources (i.e., stars and quasars; QSOs). Sufficiently bright and nearby galaxies can be easily distinguished via visual inspection of their morphology \citep{Lintott_2008} or simple heuristic cuts in the space of photometric properties \citep[e.g.,][]{Kron_1980, Leauthaud_2007}. But for the numerous objects barely above the detection limit of a survey, separating resolved sources from the unresolved remains a challenging task, and machine-learning (ML) approaches have been widely explored as a solution \citep[e.g., \citealp{SExtractor_1996,Philip_2002, Ball_2006, Henrion_2011, Vasconcellos_2011, Kovacs_2015, Miller_2017, Tachibana_2018, Stoppa_2023, Miller_2021, Beck_2022}; see][for a comprehensive review]{Sevilla-Noarbe_2018}.

Recent wide-field time-domain surveys, such as the All-Sky Automated Survey for Supernovae \citep{ASASSN_2014}, the Asteroid Terrestrial Last-Alert
System \citep{ATLAS_2011, ATLAS_2018},
the Zwicky Transient Facility \citep[ZTF;][]{ZTF_2019a, ZTF_2019b, ZTF_2020}, and BlackGEM \citep{BlackGEM_2024}, have enabled exhaustive searches of the entire transient sky, unveiling numerous Milky Way objects (e.g., cataclysmic variables and microlensing events) and extragalactic transients (e.g., supernovae, kilonovae, tidal disruption events, and exotic relativistic explosions). Millions of alerts, i.e., brightening or dimming events associated with a mixture of imaging artifacts, known variables, and (probably the most interesting) new transients, are produced overnight. For example, ZTF sends out $>$$10^5$ alert packets nightly \citep{ZTF_data_2019, ZTF_ML_2019}, and the upcoming Vera C. Rubin Observatory's Legacy Survey of Space and Time \citep[LSST;][]{LSST_2019} will generate $>$$10^7$ alerts every night. 
Automated real-time processing and candidate filtering are critical to identify young and fast evolving transients. A robust model to separate resolved--unresolved sources can support extragalactic surveys by effectively rejecting Galactic sources (i.e., stars), and vice versa. As an example, the ZTF alert stream incorporates a star--galaxy morphological score \citep[\texttt{sgscore};][\citetalias{Tachibana_2018} henceforth]{Tachibana_2018}, which proves crucial to contaminant removal in search of various extragalactic transients \citep[e.g.,][]{De_CLU_2020, BTS_I_2020, BTS_II_2020, ZTF_TDE_2021, Rehemtulla_2024}. Unfortunately, the \texttt{sgscore} model built on the Pan-STARRS1 (PS1) 3$\pi$ survey \citep{PanSTARRS_2016} does not cover declination $\le-30\degr$, and thus has limited application to surveys monitoring the southern hemisphere.

In this paper, we present our new morphological model to separate resolved--unresolved sources in the LS Data Release 10 (\dr{10}), which covers $\sim$20,000$\,\deg^2$ from the declination of $\sim$$-85\degr$ to $\sim$$+85\degr$ and outside the Galactic plane. The model is primarily designed to support the real-time search for transients in the upcoming La Silla Schmidt Southern Survey \citep{LS4_2025}. LS4 uses the ESO Schmidt telescope at La Silla Observatory, Chile, and will benefit from the complete extragalactic sky coverage of LS \dr{10}. Nevertheless, the model can be applied to many different scientific tasks other than time-domain surveys.

In Section~\ref{sec:data} we define the training set using Hubble Space Telescope (HST) observations and independent test sets for model evaluation. In Section~\ref{sec:features} we discuss the features to be included in our models, which are then trained using the Gradient Boosting algorithm \xgboost\ \citep{XGBoost_2016} described in Section~\ref{sec:model}. We evaluate the performance of our models in Section~\ref{sec:results} and show that a hybrid classifier combining two \xgboost\ models is the most robust. Finally, we show the implementation of our classifier into the LS4 pipeline in Section~\ref{sec:pipeline} and draw conclusions in Section~\ref{sec:conclusion}.

Throughout the paper, we interchangeably use the term stars to refer to unresolved, point-like sources and galaxies for resolved sources. The caveat is that active galactic nuclei (AGN), depending on their distance and brightness relative their host galaxy, can be either resolved or unresolved. For the binary classification task at hand we define stars as the positive class. In the \xgboost\ models, this corresponds to a high model score. 

We have released scripts and notebooks for all the queries and analysis to produce the results in this study at \url{https://github.com/slowdivePTG/LS-PSC}. The final catalog containing our star--galaxy scores of the entire LS \dr{10} is available at \url{https://ls-xgboost.lbl.gov/}.

\section{Model Data}\label{sec:data}
LS \dr{10} data were obtained from the Astro Data Lab \citep{Fitzpatrick_2014,Nikutta_2020}. LS catalogs two different flux measurements, the \texttt{Tractor}\footnote{\url{https://github.com/dstndstn/tractor}} flux (in \texttt{ls\_dr10.tractor}) and the aperture flux (in \texttt{ls\_dr10.apflux}). In \texttt{Tractor} each source is fit by up to 5 morphological models: point sources (PSF), round exponential galaxies (REX), de Vaucouleurs (DEV) profiles, exponential (EXP) profiles, and S\'ersic (SER) profiles. The flux from the best-fit model is then provided.\footnote{The strategy of determining the optimal morphological model is summarized at \url{https://www.legacysurvey.org/dr10/description/\#morphological-classification}.} Note that the more morphologically complicated models are not fit to sources that are well fit by simple profiles (e.g., PSF). In Section~\ref{sec:features} we will explain the \texttt{Tractor} procedure in more detail, which we use to construct features for our machine-learning model. Aperture fluxes are measured in 8 different aperture sizes in $griz$ filters: 0\farcs5, 0\farcs75, 1\farcs0, 1\farcs5, 2\farcs0, 3\farcs5, 5\farcs0, and 7\farcs0, with no curve-of-growth correction.

LS \dr{10} has a southern and a northern footprint divided at declination $=32.375\degr$. The southern footprint is covered by the 
$grz$-band observations carried out in the Dark Energy Camera Legacy Survey (DECaLS) with the Dark Energy Camera (DECam), as well as $griz$-band observations from other DECam-based programs such as the DeROSITAS survey, the BLISS+ survey, and the Dark Energy Survey (DES). In the southern footprint, $\sim$15\% sources lack $i$-band observations, while $\sim$4\%, 9\%, and 7\% of sources lack $g$, $r$, and $z$ coverage, respectively. In the northern footprint, only $grz$-band observations are available from the Beijing-Arizona Sky Survey (BASS) and the Mayall z-band Legacy Survey (MzLS), but the filter coverage is more complete, with only $\sim$0.1\%, 0.1\%, and 0.5\% of targets missing $g$, $r$, and $z$ band, respectively. Overall, $\sim$25\% of the \dr{10} sources do no have $i$-band photometry, while $\sim$15\% of the sources are not covered by at least one of the $grz$ filters. Additionally, observations carried on the southern footprint are usually deeper and have better seeing. Any model to classify LS sources must therefore be robust to missing filters and heterogeneous observing conditions.

\subsection{The HST Training Set}\label{sec:HST}
Following \citetalias{Tachibana_2018}, we construct our training set using HST observations of the 1.64\,deg$^2$ COSMOS field. \cite{Leauthaud_2007} have provided labels for HST COSMOS sources and the classification should be reliable for sources brighter than $\sim$25\,mag. We perform a spatial cross match between LS and the HST data using a 1\arcsec\ radius, which yields 428,231 unique counterparts. We exclude 155 sources with potentially problematic HST photometry (labeled as \texttt{MU\_CLASS} $=3$ by \citealp{Leauthaud_2007} suggesting artifacts) and 8 sources with an LS morphological type of DUP, which is set for Gaia sources that are coincident with an extended source -- no optical flux is provided for these sources so it is impossible to generate a valid score with our classifier. Finally we remove 187,397 sources with \texttt{MAG\_AUTO} $>25$ (Kron-like elliptical aperture magnitude in HST F814W filter), and the eventual HST training set contains 240,671 targets. This is a factor of $>$5 greater than the training set in \citetalias{Tachibana_2018}, as LS is significantly deeper than PS1.

\begin{figure*}
    \centering
    \includegraphics[width=0.83\linewidth]{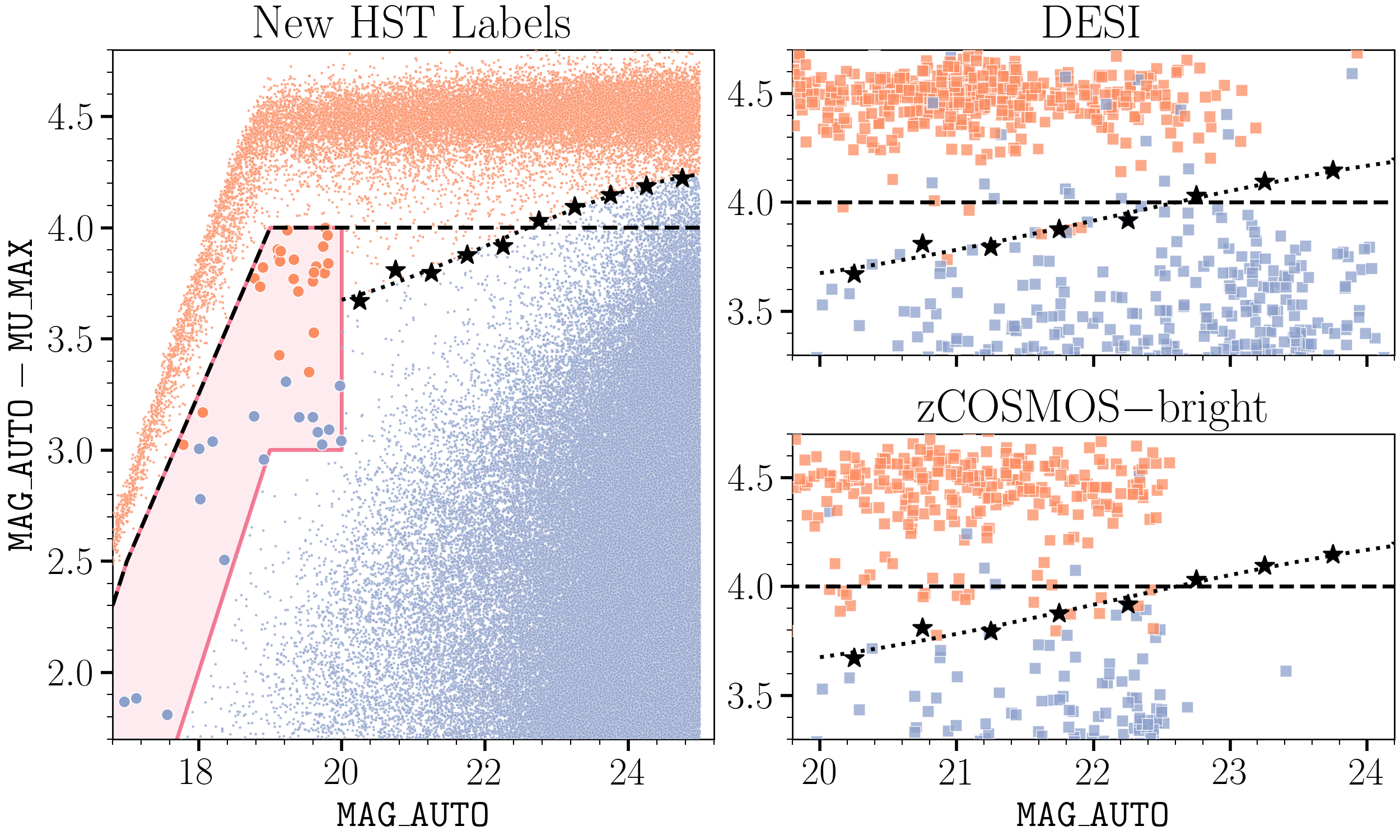}
    \caption{The modified boundary in the \magauto\ $-$ \mumax/\magauto\ plane reasonably separate stars (orange) from galaxies (blue) in the HST training set. The original boundary in \cite{Leauthaud_2007} is displayed as the dashed line. As illustrated in the text, for objects $>$20\,mag we define a new boundary as a polynomial fit (dotted line) to the boundary in each \magauto\ bin (asterisks), determined in a data-driven way. For bright objects ($<$20\,mag) with ambiguity in the \magauto\ $-$ \mumax/\magauto\ plane (Equation~(\ref{eq:VI}); red shaded region), we determine their labels by visual inspection. {\it Left:} the labels adopted in the training set. {\it Right:} the subset of training-set targets with a spectroscopic classification by DESI \dr{1} (top) and zCOSMOS-bright (bottom) near the boundary, color-coded by their spectroscopic classes. Stars below the dotted line and galaxies above the dotted line are mislabeled targets in our training set.}
    \label{fig:relabel}
\end{figure*}

While \citet{Leauthaud_2007} have demonstrated that HST point sources and resolved sources are well separated in the flux/surface-brightness space (see their Figure~5), their adopted heuristic boundary was drawn to prioritize the purity of the galaxy catalog to enable weak gravitational lensing studies. However, in the workflow searching for extragalactic transients, the completeness of the galaxy sample is arguably more crucial than its purity. In addition, as we will show later, for bright objects our machine learning models produce near-perfect class separation, meaning label noise can dominate ``misclassification'' during model evaluation. For these reasons, we carefully revisit the labels in the training set.

Intuitively, the surface brightness of unresolved, point sources is proportional to the overall flux, whereas for resolved sources the surface brightness does not necessarily correlate. It has been found in \cite{Leauthaud_2007} that for faint targets ($\gtrsim$20\,mag) in their sample, the boundary between stars and galaxies is roughly
\begin{equation*}
\texttt{MAG\_AUTO} - \texttt{MU\_MAX} = 4,
\end{equation*}
where \magauto\ and \mumax\ (peak surface brightness above background) are outputs from \texttt{Source Extractor} \citep{SExtractor_1996}. Stars are clustered around $\texttt{MAG\_AUTO} - \texttt{MU\_MAX} \simeq 4.5$, while galaxies span a wide range in $\texttt{MAG\_AUTO} - \texttt{MU\_MAX}$ below the cut. We adopt the same formalism but define a boundary that is brightness-dependent,
\begin{equation}\label{eq:bd}
    \texttt{MAG\_AUTO} - \texttt{MU\_MAX} = P(\texttt{MAG\_AUTO}),
\end{equation}
where $P$ is a polynomial. To determine the coefficients, we first evaluate the boundaries in 10 \magauto\ bins between 20 and 25. In each bin, we calculate the 1-dimensional Gaussian kernel density estimate (KDE) of the probability density function (PDF) of \magauto\ $-$ \mumax, with a kernel size of 0.075. In the derived PDF we find that the distribution of galaxies has a long tail towards the stars, whereas the stars are constrained in a narrow and sharp peak. To achieve a highly complete selection of galaxies, we define the boundary at which the PDF starts to rise at the edge of the cluster of stars, or quantitatively, when the slope of the PDF (derivative with respect to \magauto\ $-$ \mumax) exceeds 0.02. In Figure~\ref{fig:relabel} we show the boundaries in each bin as asterisks, which can be well fit by a 3rd-order polynomial (black dotted line). The new boundary $P(\texttt{MAG\_AUTO})$ is clearly not a constant as the original cut (black dashed line), but instead increases with \magauto. For brighter targets, we cannot obtain a valid PDF estimate due to the scarcity of data within the gap between the two classes. Instead, we manually relabel all 66 targets in the gap (red shaded region in Figure~\ref{fig:relabel}), defined as
\begin{equation}\label{eq:VI}
    \left.\begin{array}{cc}
         \mathrm{(i)} & \texttt{MU\_CLASS} = 1,\\
         \mathrm{(ii)} & \texttt{MAG\_AUTO} < 20,\\
          & \texttt{MU\_MAX} < 16.5,\\
         \mathrm{(iii)} & \mathrm{or} \\
          & \texttt{MAG\_AUTO - MU\_MAX} > 3.
    \end{array}\right\}
\end{equation}
In \cite{Leauthaud_2007} they have all been labeled as $\texttt{MU\_CLASS} = 1$, i.e., resolved sources. We visually inspect their LS \dr{10} images on the LS Sky Viewer\footnote{\url{https://www.legacysurvey.org/viewer}} to check if there is apparent galaxy structure. They are also matched to a couple of spectroscopic catalogs. Most of them were cataloged in the DESI Data Release 1 \citep[\dr{1};][]{DESI_DR1_2025}, the Sloan Digital Sky Survey \citep{SDSS_2000, SDSS_2022}, and/or zCOSMOS \citep{zCOSMOS_2007, zCOSMOS_2009}, of which the spectroscopic types are adopted as their new labels. There are 10 objects that are not included in any of the catalogs. Apart from one object (\texttt{ls\_id} $=10995472384671213$) which is obviously a star blended with a background galaxy, all sources appear point-like and isolated. One of them was observed in a spectroscopic survey of red galaxies \citep{Ross_2008} but turned out to be a star. All other objects show significant proper motions measured by Gaia and should also be foreground stars. After this process, 24 out of 66 objects are relabeled as stars, shown as the orange dots on the left panel of Figure~\ref{fig:relabel}. Around 20\,mag, both the visual inspection results and the KDE method suggest a new boundary at \magauto\ $-$ \mumax\ $\simeq3.5$, inconsistent with the original boundary at 4.

We further evaluate the new labels with two spectroscopic samples -- the DESI \dr{1}, and zCOSMOS 10k-Bright Spectroscopic Sample \citep[zCOSMOS-bright;][]{zCOSMOS_2009}. Both datasets offer spectroscopic classifications for a subset of targets in our training set. We apply 1\arcsec\ spatial cross matches between both datasets and the training set, and display the matches as well as their labels in Figure~\ref{fig:relabel}. For DESI targets we simply adopt their spectroscopic classes, whereas sources labeled as QSOs are not displayed. zCOSMOS-bright does not explicitly provide spectroscopic classes, yet a spectroscopic redshift is evaluated for each target which is 0 for stars. We include targets with a very secure redshift measurement that is not a broad-line AGN (confidence number $=$ 3 or 4 and decimal place modifiers $>$ 0.1, or confidence number $=$ 9 and decimal place modifiers $=$ 0.3 or 0.5).\footnote{The zCOSMOS Confidence Class rules are explained in detail at \url{https://vizier.cds.unistra.fr/viz-bin/VizieR-n?-source=METAnot&catid=21840218&notid=1&-out=text}.} There is one object with a non-zero $z=0.0004$ (so by definition a galaxy) but appears unresolved on LS imaging and exhibits Balmer absorption at $z=0$ in its DESI spectrum -- we therefore remove this object as it is most likely a star.

For both samples, the new boundary defined by Equation~(\ref{eq:bd}) separates the two classes reasonably well, especially for zCOSMOS-bright targets. We note that stars $\gtrsim$19\,mag are strongly under-represented in the DESI \dr{1} due to a targeting strategy focused on luminous red galaxies \citep[LRGs;][]{DESI_LRG_2023}, emission line galaxies \citep[ELGs;][]{DESI_ELG_2023}, and QSOs \citep{DESI_QSO_2023}. As a result, there are few DESI stars near the new boundary. Nevertheless, our new boundary better handles edge cases than the linear cut adopted in \cite{Leauthaud_2007}, and we shall utilize the new labels when training and evaluating our models.

\begin{figure*}
    \centering
    \includegraphics[width=\linewidth]{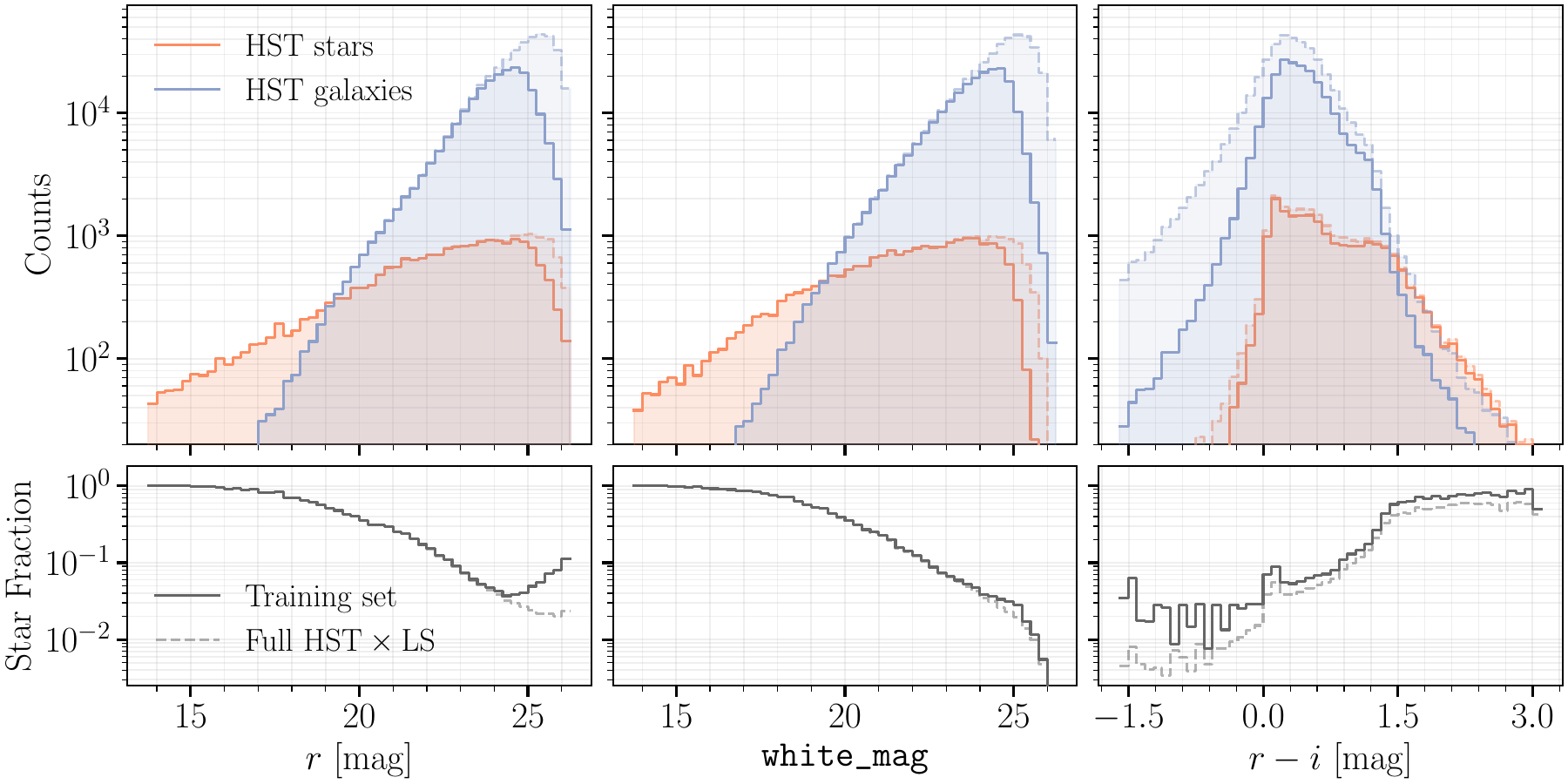}
    \caption{Distributions of stars and galaxies suggest that, in the faintest magnitude bins with single band filters (e.g., $r$), the star fraction deviates significantly from the reality. Using the flux-weighted average magnitude \texttt{white\_mag} (defined in the text) alleviates the bias.
    {\it Upper:} the distribution of HST stars and galaxies in the COSMOS field among all cross-matches between LS and HST (dashed line) as well as in the training set which excludes targets fainter than \magauto\ $=25$ (solid line). {\it Bottom:} the corresponding stellar fraction. The distributions are shown in terms of the $r$-band magnitude, \texttt{white\_mag}, and the $r-i$ color.}
    \label{fig:bias}
\end{figure*}

Finally we note that the HST$\times$LS dataset is highly imbalanced. On its bright end, the dataset consists of mostly unresolved stars, but on the faint end it is fully dominated by galaxies. Nevertheless, we expect the HST$\times$LS dataset exhibits representative source distributions of what is observed at high Galactic latitudes in the entire LS dataset. When defining the training set, we exclude targets $>$25\,mag in the HST F814W filter, whose central wavelength is close to that of the LS $i$ filter (784\,nm). Here we briefly discuss the bias in source distributions introduced by the magnitude cut in such a red filter. 

Figure~\ref{fig:bias} shows the distribution of the star--galaxy counts and the star fraction over a range of magnitude and color for both the full HST$\times$LS dataset containing all the cross-matched results (dashed lines), as well as the training set when a magnitude cut is applied (solid lines). The star--galaxy separation for faint objects is less confident. In \cite{Leauthaud_2007}, aside from the $\texttt{MAG\_AUTO} - \texttt{MU\_MAX}=4$ cut, another boundary between the resolved--unresolved sources is $\texttt{MU\_MAX} = 21.5$. We adopt this $\texttt{MU\_MAX}$ boundary, but define $\texttt{MAG\_AUTO} - \texttt{MU\_MAX}=P(\magauto=25)\simeq 4.2$ as the new boundary in consistency with the brighter sources. Note that this separation is simply for illustration purpose. In the full HST$\times$LS dataset, the distributions of both stars and galaxies do not deviate significantly from a power-law until $r\simeq25$, suggesting that both the star and galaxy samples are nearly complete before hitting the detection limit of LS. In the training set, the turning points of the distributions pop up earlier, especially for galaxies (left panel of Figure~\ref{fig:bias}). In the faintest magnitude bins where most of our targets lie, the stars are overrepresented and the their number fraction starts to rise among objects $r\gtrsim24$\,mag.

Stars and galaxies appear fairly different in broad-band colors, as displayed on the right panel of Figure~\ref{fig:bias}. On average stars observed in the COSMOS field are redder than galaxies. While the bulk majority of stars and galaxies show a $r-i$ color between $\sim$0--1.5\,mag, the reddest objects ($r-i>1.5$\,mag) consist mostly of late-type stars, whereas the bluest objects ($r-i<0$\,mag) are predominantly star-forming galaxies. By making a cut in $\sim$$i$-band flux, blue sources will be preferentially excluded in the dataset compared to sources with the same total flux over all filters but a much redder color. Thus, we inevitably exclude a large fraction of star-forming galaxies and introduce an overabundance of stars in the faintest magnitude bins. 

\citetalias{Tachibana_2018} defined the ``white'' flux of sources, which is an weighted average of the flux in all available filters $f$ ($griz$ for LS),
\begin{equation}\label{eq:white_flux}
    \texttt{white\_flux} = \frac{\sum w_f\times \texttt{flux}\_f}{\sum w_f},
\end{equation}
where the weight $w_f$ is the S/N squared, $w_f=(\texttt{snr\_}f)^2$. For non-detection, $w_f$ is set to 0. The corresponding magnitude \texttt{white\_mag} better characterizes the overall brightness of sources with either missing filters or extreme colors, as opposed to magnitude in individual filters. In the middle panel of Figure~\ref{fig:bias} we also show the distribution of stars and galaxies in terms of \texttt{white\_mag}. While there is a bump in the star fraction at $\sim$25\,mag in the training set, the over-representation of stars is alleviated in individual \texttt{white\_mag} bins. Later, when evaluating and comparing the performance of different models on sources, \texttt{white\_mag} will be used to represent their brightness.

\subsection{Test Sets}
We test our models on two independent datasets from wide-field spectroscopic surveys, DESI \dr{1} \citep{DESI_DR1_2025} and SDSS \dr{17} \citep{SDSS_2022}. Both surveys provide spectroscopic classification (star, galaxy, or QSO\footnote{Traditionally QSOs refer to bright AGNs as point sources, yet in DESI and SDSS the ``QSO'' class is more inclusive and some lower luminosity AGNs also fall into the category.}) for sources spanning a significant range of declination, with which we test the generalization ability of our models. Additionally, we construct another dataset of bright stars using Gaia \dr{3} \citep{GaiaDR3_2023}, which has the most complete sky coverage, to test the robustness of our models in identifying stars when some filters are missing.

\subsubsection{DESI}\label{sec:DESI}
There are 20,462,470 unique matches of DESI \dr{1} objects within 1\arcsec\ of LS \dr{10} sources with useful spectra ($\texttt{zwarn}=0$, meaning no errors due to instrument or data and no flags from spectral fitting issues). Occasionally individual LS sources may be matched to multiple DESI records, indicating that multiple spectra were obtained in the different survey types (e.g., primary or secondary). If the DESI spectroscopic classifications agree with each other, we adopt them as labels. Sources with inconsistent DESI classifications are dropped, leaving 20,419,060 sources. We further remove 18,036 targets in the COSMOS field, i.e., the training set. Then we add a couple of purity filters to reduce the labeling noise by excluding
\begin{itemize}
    \item 1,550,650 sources with ambiguous spectroscopic template fitting: $\texttt{deltachi2} < 50$.
    \item 1,078,906 sources in tiles with low effective exposure time: $\texttt{efftime\_spec} < 100$, which is insufficient for valid spectroscopic modeling even for many bright objects.
    \item DESI galaxies with (i) $\texttt{z}<0.001$ (62,249), which are mostly misclassified stars, or (ii) $\texttt{z}>1.6$ (78,873), which are susceptible to nonphysical fits due to lack of prominent emission lines.
    \item 1,640,726 DESI QSOs and 644,061 targets in the dedicated QSO survey that are not classified as QSOs. QSOs, depending on their brightness relative to the host galaxies, could be either resolved or unresolved. We further remove targets in the QSO survey because a non-negligible fraction of broad-line QSOs embedded in ELGs are misclassified as galaxies by the standard \texttt{Redrock} pipeline \citep{DESI_QSO_VI_2023}.
    \item 72,807 potential artifacts near bright stars. Sometimes the profile of a bright star may slightly deviate from a PSF (usually due to saturation). Then after a PSF model is fit and extracted, the residual flux could be extracted as a new source, which often has an extended profile. They usually lie within $\sim$1\arcsec\ from a bright star but are mostly artifacts. We use Gaia \dr{3}, which has a nice coverage of the bright stars in the LS footprint, to identify these artifacts -- when multiple LS/DESI sources are found within 1\arcsec\ to a Gaia star candidate (defined in Section~\ref{sec:Gaia}), we only keep the sources with an almost exact match\footnote{For bright Gaia sources, the LS pipeline would try to extract the source at the same location in the LS images, so the coordinates should be almost the same.} (in practice, separation no larger than a milliarcsecond).
    In Section~\ref{sec:pipeline} we will show how we handle these artifacts in our final data products.
\end{itemize}
Note that different filters overlap. Finally we remove 60,748 objects with no valid detections in any filters ($\texttt{snr\_}f \le 0$, or no flux measurements in any apertures; usually because of saturation). The ultimate DESI test set consists of 15,354,071 targets.

\subsubsection{SDSS}\label{sec:SDSS}
The construction of the SDSS test set is very similar to that of the DESI test set. There are 4,170,260 unique matches of SDSS \dr{17} objects within 1\arcsec\ of LS \dr{10} sources with $\texttt{zwarn}=0$. We exclude the 680 targets in the COSMOS field. Then we also remove 1,404 SDSS galaxies with $\texttt{z}<0.001$, 747,125 SDSS QSOs, and 3,207 potential artifacts near bright Gaia stars for the same reasons. Eventually by excluding 60,704 sources no valid LS photometry, we obtain a sample of 3,357,255 targets.

There are a small number of bright and extended galaxies misclassified as stars by SDSS. We cross matched the SDSS catalog with the Siena Galaxy Atlas 2020 (SGA) catalog \citep{SGA_2023}, and found 270 SDSS targets within 1\farcs5 of an extended SGA galaxy labeled as stars, for which we flip their classes.

\subsubsection{Gaia}\label{sec:Gaia}
There are 235,243,762 unique matches in Gaia \dr{3} within 1\arcsec\ of LS \dr{10} sources, from which we aim to construct a subset of stars with high purity. Following the approach in \citetalias{Tachibana_2018}, we require high-significance detection in either the parallax $\varpi$ or the proper motion $\mu$. The significance of $\varpi$ is defined as $\varpi/\sigma_\varpi$, which is called {\tt parallax\_over\_error} in the Gaia database. The total proper motion is a combination of the proper motions in Right Ascension $\mu_{\alpha*}$ and Declination $\mu_\delta$ ({\tt pmra} and {\tt pmdec} in the database),
\begin{equation*}
    \mu^2 = \mu_{\alpha*}^2 + \mu_\delta^2.
\end{equation*}
The uncertainty in the total proper motion $\sigma_\mu$ can be calculated with
\begin{equation*}
    \sigma^2 = \frac{\mu_{\alpha*}^2}{\mu^2}\sigma_{\mu_{\alpha*}}^2 + \frac{\mu_{\delta}^2}{\mu^2}\sigma_{\mu_{\delta}}^2 + 2\frac{\mu_{\alpha*}\mu_\delta}{\mu^2}\rho(\mu_{\alpha*}\mu_\delta)\sigma_{\mu_{\alpha*}}\sigma_{\mu_{\delta}},
\end{equation*}
where $\rho(\mu_{\alpha*}\mu_\delta)$ ({\tt pmra\_pmdec\_corr} in the database) is the correlation coefficient between $\mu_{\alpha*}$ and $\mu_\delta$. We require either the parallax significance or total proper motion significance $\ge$5, removing 32,277,565 sources. 
We further require the sources to pass the filter defined by Equation (C.2) in \cite{Lindegren_2018} to eliminate 24,220,059 targets with potential issues in BP and RP photometry {due to crowding, most of which are faint sources in high stellar density fields \citep{Gaia_2018}. By cross matching sources with a significant parallax or proper motion with DESI, we find that $\sim$5\% of sources failing (C.2) are classified as galaxies, whereas galaxy contamination is negligible among those passing (C.2).} Again, we remove 292,486 sources with no valid detection in any filter, leaving 178,453,652 sources matched with reliable Gaia stars. Finally, we remove 1,971,413 potential artifacts near bright Gaia stars. The final Gaia test set contains 176,482,239 sources.

\section{Model Features}\label{sec:features}
LS provides the morphological typing by fitting each source to multiple morphological profiles (PSF, REX, DEV, EXP, SER) and finding the optimal model. Sources that are best fit to a PSF model are considered as unresolved, point-like objects, whereas resolved, extended sources should be better fit to other morphological models. Throughout the paper we will compare the performance of our machine learning model to the LS morphological typing. 
The goodness of fit is quantified using \texttt{dchisq}, the difference of $\chi^2$ when fitting a source to a given model versus the $\chi^2$ assuming there is no source at the sky location. For each source, five values (\texttt{dchisq\_1} to \texttt{dchisq\_5} in the LS database) are recorded, corresponding to fits with the five models. Positive \texttt{dchisq} indicates better fits, and we define the first set of features as the differences in {\tt dchisq} by adopting profiles of extended sources versus the PSF model ($\textit{gal}=$ 2--5, corresponding to REX, DEV, EXP, and SER)
\begin{equation}\label{eq:feat_dchisq}
    \texttt{delta\_dchisq}(\text{gal}) = \frac{\texttt{dchisq\_}{\text{gal}} - \texttt{dchisq\_1}}{\texttt{snr}^2}.
\end{equation}
The improvement in {\tt dchisq} is normalized by the average \texttt{snr}$^2$, defined as the quadratic mean of the signal to noise ratio (S/N) in all 4 filters,
\begin{equation}
    \texttt{snr}^2 = \sum_{f=\texttt{griz}} \texttt{snr\_}{f}^2/4,
\end{equation}
where any $\texttt{snr\_}f\le 0$ is set to 0.\footnote{LS allows negative fluxes for faint targets, but here we still treat them as non-detections.} We try to eliminate any explicit information of the source S/N in constructing model features, to ensure the generalization ability of our models for sources covering a wide range of brightness. 

We note that LS does not always fit sources to all 5 morphological models. In general, a source is first fit to the simplest PSF and REX profiles. Only when the REX model is strongly preferred will the source be fit to more complex DEV or EXP profiles. Similarly, only when DEV and EXP are preferred (increasing \texttt{dchisq} by at least 9 compared to the simple PSF model) will the source be fit to the SER profile with the highest complexity. When $\texttt{dchisq\_3,4,5}=0$ (suggesting the certain model is not fit), we set the corresponding features to NaN to indicate the values are missing.

There are also a non-negligible number of sources that are fit to only the PSF model ($\texttt{dchisq\_2}=0$) or not fit at all ($\texttt{dchisq\_1,2}=0$). In the first case, the targets are forced to be point sources, which are also indicated in  \texttt{fitbits},\footnote{Detailed descriptions of \texttt{fitbits} can be found at \url{https://www.legacysurvey.org/dr10/bitmasks/}.} a bit-mask detailing any peculiarities regarding how a source was fit, on bit 0 (\texttt{FORCED\_POINTSOURCE}) or 12 (\texttt{GAIA\_POINTSOURCE}). On the DESI test set, we find $\sim$10\% of targets are forced to be point sources, among which $\sim$99.5\% are bright stars. In the second case, the targets are not fit to any models, which usually indicates coincidence with bright extended galaxies in the SGA catalog. On the DESI test set, only $\sim$0.6\% of targets are not fit to any models, of which $\sim$99.5\% are galaxies. While these two groups of targets show distinct natures, by definition their $\texttt{delta\_dchisq}(gal)$ will all be set to NaN and become indistinguishable. We thus introduce a new feature
\begin{equation}\label{eq:feat_mask}
    \texttt{mask} = \left\{
    \begin{array}{cl}
        0, & \mathrm{Fit\ to\ both\ PSF\ and\ REX};\\
        1, & \mathrm{Not\ fit}; \\
        2, & \mathrm{Forced\ to\ be\ PSF}.
    \end{array}
    \right.
\end{equation}
Specifically, we set \texttt{mask} $=1$ when both 
\begin{equation*}
\texttt{dchisq\_1}=\texttt{dchisq\_}\textit{gal}=0
\end{equation*}
and
\begin{equation*}
\texttt{fitbits}\ \&\ (2^0 + 2^{12})=0,
\end{equation*}
and set \texttt{mask} $=2$ when
\begin{equation*}
\texttt{dchisq\_1}>0,\ \texttt{dchisq\_}\textit{gal}=0
\end{equation*}
and
\begin{equation*}
\texttt{fitbits}\ \&\ (2^0 + 2^{12})\neq 0.
\end{equation*}
In all other cases, \texttt{mask} $=0$. We note that a very small fraction ($\ll$0.1\%) of DESI targets show inconsistent \texttt{dchisq} and \texttt{fitbits} values, e.g., some forced point sources indicated by \texttt{fitbits} show \texttt{dchisq\_1} $=0$. These targets are extremely rare, with no strong preference for stars or galaxies, so we also set their \texttt{mask} to 0.

Additionally, we also construct features from the aperture flux measurements. To avoid features that are explicitly associated with the brightness of sources, i.e., the absolute flux measurements in each aperture, the proposed features are based on the flux ratios between the 7 pairs of consecutive apertures, which only contain information of the radial flux distribution,
\begin{equation}\label{eq:feat_ap}
    \texttt{apratio\_}f(\texttt{size}(k), \texttt{size}(k+1)) = \frac{\texttt{apflux\_}f\texttt{\_}k}{\texttt{apflux\_}f\texttt{\_}k+1},
\end{equation}
where $k=1$--7, corresponding to the 7 aperture sizes. 
If \texttt{apflux}$\_f\_k+1$ = 0, the corresponding \texttt{apratio}$\_f$ is set to NaN. We expect the flux of unresolved point sources to be more concentrated within smaller apertures (depending on the average seeing over the stacked images), and the unresolved sources show more extended flux distributions.

Some objects are missing measurements in multiple filters, so we again define the ``white'' aperture flux as the weighted average of the flux measured in all available filters $f$ among $griz$
\begin{equation}\label{eq:apflux_white}
    \texttt{apflux\_white\_}k = \frac{\sum w_f(k) \times \texttt{apflux\_}f\_k}{\sum w_f(k)},
\end{equation}
where the weight $w_f(k)$ is again the S/N of the aperture flux squared 
\begin{equation*}
    w_f(k) = (\texttt{apflux\_}f\_k)^2\times \texttt{apflux\_ivar\_}f\_k.
\end{equation*} 
If the source is not observed in filter $f$ or if certain apertures are masked by LS (usually due to saturation), the corresponding \texttt{apflux\_ivar\_}$f\_k$ is set to 0. With the white fluxes we construct the features \texttt{Feat[ap\_white]} using Equation~(\ref{eq:feat_ap}).
Our fiducial model includes \texttt{delta\_dchisq}(gal), \texttt{mask}, and \texttt{apratio\_white}, so we call it the White model. 

As a comparison, we construct another model containing flux ratios in individual filters, i.e., \texttt{apratio\_g}, \texttt{apratio\_r}, \texttt{apratio\_z} instead of \texttt{apratio\_white}.
The corresponding feature is set to NaN if the $(k+1)$-th aperture is masked or the S/N of the flux measurement is below 3.
Note that only filters $grz$ are included in the model due to the very incomplete sky coverage in $i$. As we will show in Section~\ref{sec:results}, this $grz$ model significantly outperforms the fiducial White model when all 3 filters are available. However, when even a single filter is missing, the data input becomes very sparse and the model performance significantly degrades.

To feed our model with knowledge of individual filters while retaining the robustness of the White model with missing filters, we design a Hybrid model, whose outcome is a linear combination of the predictions from two independently trained \xgboost\ models. Both sub-models still contain \texttt{delta\_dchisq}(gal) and \texttt{mask}, and as for \texttt{apratio}, the first sub-model adopts the weighted average of fluxes in $g$ and $r$ (\texttt{apratio\_gr}), whereas the second one adopts fluxes in $i$ and $z$ (\texttt{apratio\_iz}),
\begin{equation}\label{eq:apflux_hybrid}
\begin{aligned}
    \texttt{apflux\_gr}\_k &= \frac{w_g(k) \ \texttt{apflux\_g}\_k + w_r(k) \ \texttt{apflux\_r}\_k}{w_g(k) + w_r(k)},\\
    \texttt{apflux\_iz}\_k &= \frac{w_i(k) \ \texttt{apflux\_i}\_k + w_z(k) \ \texttt{apflux\_z}\_k}{w_i(k) + w_z(k)},
\end{aligned}
\end{equation}
which are set to NaN if both $w_f=0$. In this way, each sub-model extracts information from an broad blue ($gr$) or red ($iz$) filter effectively. With these ``broad-band'' aperture fluxes, we define the flux ratios \texttt{apratio\_gr} and \texttt{apratio\_iz} using Equation~(\ref{eq:feat_ap}). Eventually in making predictions, each sub-model produces a score, $\mathrm{score}_{gr}$ and $\mathrm{score}_{iz}$, and the final output is the weighted average of the two
\begin{equation}\label{eq:score_hybrid}
    \mathrm{score} = \frac{\alpha_\mathrm{hyb}\mathrm{det}_{gr}\times\mathrm{score}_{gr} + (1 - \alpha_\mathrm{hyb})\mathrm{det}_{iz}\times\mathrm{score}_{iz}}{\alpha_\mathrm{hyb}\mathrm{det}_{gr} + (1 - \alpha_\mathrm{hyb})\mathrm{det}_{iz}}.
\end{equation}
Here $\alpha_\mathrm{hyb}\in(0,1)$ is a parameter of the relative importance between the two sub-models to be determined in the training. The weights $\mathrm{det}_{gr}, \mathrm{det}_{iz}\in\{0, 1\}$ indicate whether both filters adopted in each sub-model are missing or not. Following this definition, even if a source is observed in only 1 filter, e.g., $z$, then \texttt{apratio\_iz} would still have meaningful values and the model would simply present $\mathrm{score}_{iz}$ as the output.

Features included in each \xgboost\ model are summarized in Table~\ref{tab:features}. We will show in Section~\ref{sec:results} that when there are few missing values in the input, the $grz$ model performs the best in separating resolved sources from the unresolved. However, in a more realistic dataset where a non-negligible fraction of sources are not observed in a few filters, the $grz$ model dramatically downgrades and the Hybrid model performs significantly better, so we will use the Hybrid model to produce the LS resolved--unresolved source catalog.

Finally we note that, unlike many other surveys such as PS1, LS does not provide either PSF or \cite{Kron_1980} photometry, so we cannot build features based on the ratio of fluxes measured from different photometric methods, demonstrated to be useful in star--galaxy separation \citepalias{Tachibana_2018}. LS also presents parameters related to the size and shape of the source (e.g., half-light radius and ellipticity), but as they are only calculated for sources which are best fit to an extended profile, we exclude them from our models.

\begin{deluxetable*}{lll}\label{tab:features}
\tabletypesize{\footnotesize}
\tablehead{
    \colhead{Name} & \colhead{Definition} & \colhead{Description}
}
\tablecaption{A summary of features included in each \xgboost\ model.}
\startdata
\multicolumn{3}{l}{Common features included in all models (5)} \\
\hline
\texttt{delta\_dchisq}(REX) & (\ref{eq:feat_dchisq}) & Improvement in fitting the source by adopting the REX profile compared to the PSF profile\\
\texttt{delta\_dchisq}(DEV) & (\ref{eq:feat_dchisq}) & Improvement in fitting the source by adopting the DEV profile compared to the PSF profile\\
\texttt{delta\_dchisq}(EXP) & (\ref{eq:feat_dchisq}) & Improvement in fitting the source by adopting the EXP profile compared to the PSF profile\\
\texttt{delta\_dchisq}(SER) & (\ref{eq:feat_dchisq}) & Improvement in fitting the source by adopting the SER profile compared to the PSF profile\\
\texttt{mask} & (\ref{eq:feat_mask}) & A flag indicating whether the source is not fit to any profiles or to the PSF profile only \\
\hline
\multicolumn{3}{l}{Unique features in the White model (7)} \\
\hline
\texttt{apratio\_white}(0\farcs5, 0\farcs75) & (\ref{eq:feat_ap}), (\ref{eq:apflux_white}) & The ratio of white flux in the 0\farcs5 and 0\farcs75 apertures \\
\texttt{apratio\_white}(0\farcs75, 1\farcs0) & (\ref{eq:feat_ap}), (\ref{eq:apflux_white}) & The ratio of white flux in the 0\farcs75 and 1\farcs0 apertures \\
\texttt{apratio\_white}(1\farcs0, 1\farcs5) & (\ref{eq:feat_ap}), (\ref{eq:apflux_white}) & The ratio of white flux in the 1\farcs0 and 1\farcs5 apertures \\
\texttt{apratio\_white}(1\farcs5, 2\farcs0) & (\ref{eq:feat_ap}), (\ref{eq:apflux_white}) & The ratio of white flux in the 1\farcs5 and 2\farcs0 apertures \\
\texttt{apratio\_white}(2\farcs0, 3\farcs5) & (\ref{eq:feat_ap}), (\ref{eq:apflux_white}) & The ratio of white flux in the 2\farcs0 and 3\farcs5 apertures \\
\texttt{apratio\_white}(3\farcs5, 5\farcs0) & (\ref{eq:feat_ap}), (\ref{eq:apflux_white}) & The ratio of white flux in the 3\farcs5 and 5\farcs0 apertures \\
\texttt{apratio\_white}(5\farcs0, 7\farcs0) & (\ref{eq:feat_ap}), (\ref{eq:apflux_white}) & The ratio of white flux in the 5\farcs0 and 7\farcs0 apertures \\
\hline
\multicolumn{3}{l}{Unique features in the $grz$ model (21)} \\
\hline
\texttt{apratio\_g}(0\farcs5, 0\farcs75) & (\ref{eq:feat_ap}) & The ratio of $g$-band flux in the 0\farcs5 and 0\farcs75 apertures \\
\texttt{apratio\_g}(0\farcs75, 1\farcs0) & (\ref{eq:feat_ap}) & The ratio of $g$-band flux in the 0\farcs75 and 1\farcs0 apertures \\
\texttt{apratio\_g}(1\farcs0, 1\farcs5) & (\ref{eq:feat_ap}) & The ratio of $g$-band flux in the 1\farcs0 and 1\farcs5 apertures \\
\texttt{apratio\_g}(1\farcs5, 2\farcs0) & (\ref{eq:feat_ap}) & The ratio of $g$-band flux in the 1\farcs5 and 2\farcs0 apertures \\
\texttt{apratio\_g}(2\farcs0, 3\farcs5) & (\ref{eq:feat_ap}) & The ratio of $g$-band flux in the 2\farcs0 and 3\farcs5 apertures \\
\texttt{apratio\_g}(3\farcs5, 5\farcs0) & (\ref{eq:feat_ap}) & The ratio of $g$-band flux in the 3\farcs5 and 5\farcs0 apertures \\
\texttt{apratio\_g}(5\farcs0, 7\farcs0) & (\ref{eq:feat_ap}) & The ratio of $g$-band flux in the 5\farcs0 and 7\farcs0 apertures \\
\texttt{apratio\_r}(0\farcs5, 0\farcs75) & (\ref{eq:feat_ap}) & The ratio of $r$-band flux in the 0\farcs5 and 0\farcs75 apertures \\
\texttt{apratio\_r}(0\farcs75, 1\farcs0) & (\ref{eq:feat_ap}) & The ratio of $r$-band flux in the 0\farcs75 and 1\farcs0 apertures \\
\texttt{apratio\_r}(1\farcs0, 1\farcs5) & (\ref{eq:feat_ap}) & The ratio of $r$-band flux in the 1\farcs0 and 1\farcs5 apertures \\
\texttt{apratio\_r}(1\farcs5, 2\farcs0) & (\ref{eq:feat_ap}) & The ratio of $r$-band flux in the 1\farcs5 and 2\farcs0 apertures \\
\texttt{apratio\_r}(2\farcs0, 3\farcs5) & (\ref{eq:feat_ap}) & The ratio of $r$-band flux in the 2\farcs0 and 3\farcs5 apertures \\
\texttt{apratio\_r}(3\farcs5, 5\farcs0) & (\ref{eq:feat_ap}) & The ratio of $r$-band flux in the 3\farcs5 and 5\farcs0 apertures \\
\texttt{apratio\_r}(5\farcs0, 7\farcs0) & (\ref{eq:feat_ap}) & The ratio of $r$-band flux in the 5\farcs0 and 7\farcs0 apertures \\
\texttt{apratio\_z}(0\farcs5, 0\farcs75) & (\ref{eq:feat_ap}) & The ratio of $z$-band flux in the 0\farcs5 and 0\farcs75 apertures \\
\texttt{apratio\_z}(0\farcs75, 1\farcs0) & (\ref{eq:feat_ap}) & The ratio of $z$-band flux in the 0\farcs75 and 1\farcs0 apertures \\
\texttt{apratio\_z}(1\farcs0, 1\farcs5) & (\ref{eq:feat_ap}) & The ratio of $z$-band flux in the 1\farcs0 and 1\farcs5 apertures \\
\texttt{apratio\_z}(1\farcs5, 2\farcs0) & (\ref{eq:feat_ap}) & The ratio of $z$-band flux in the 1\farcs5 and 2\farcs0 apertures \\
\texttt{apratio\_z}(2\farcs0, 3\farcs5) & (\ref{eq:feat_ap}) & The ratio of $z$-band flux in the 2\farcs0 and 3\farcs5 apertures \\
\texttt{apratio\_z}(3\farcs5, 5\farcs0) & (\ref{eq:feat_ap}) & The ratio of $z$-band flux in the 3\farcs5 and 5\farcs0 apertures \\
\texttt{apratio\_z}(5\farcs0, 7\farcs0) & (\ref{eq:feat_ap}) & The ratio of $z$-band flux in the 5\farcs0 and 7\farcs0 apertures \\
\hline
\multicolumn{3}{l}{Unique features in the Hybrid model (14)} \\
\hline
\texttt{apratio\_gr}(0\farcs5, 0\farcs75) & (\ref{eq:feat_ap}), (\ref{eq:apflux_hybrid}) & The ratio of the weighted average flux of $g$ and $r$ in the 0\farcs5 and 0\farcs75 apertures \\
\texttt{apratio\_gr}(0\farcs75, 1\farcs0) & (\ref{eq:feat_ap}), (\ref{eq:apflux_hybrid}) & The ratio of the weighted average flux of $g$ and $r$ in the 0\farcs75 and 1\farcs0 apertures \\
\texttt{apratio\_gr}(1\farcs0, 1\farcs5) & (\ref{eq:feat_ap}), (\ref{eq:apflux_hybrid}) & The ratio of the weighted average flux of $g$ and $r$ in the 1\farcs0 and 1\farcs5 apertures \\
\texttt{apratio\_gr}(1\farcs5, 2\farcs0) & (\ref{eq:feat_ap}), (\ref{eq:apflux_hybrid}) & The ratio of the weighted average flux of $g$ and $r$ in the 1\farcs5 and 2\farcs0 apertures \\
\texttt{apratio\_gr}(2\farcs0, 3\farcs5) & (\ref{eq:feat_ap}), (\ref{eq:apflux_hybrid}) & The ratio of the weighted average flux of $g$ and $r$ in the 2\farcs0 and 3\farcs5 apertures \\
\texttt{apratio\_gr}(3\farcs5, 5\farcs0) & (\ref{eq:feat_ap}), (\ref{eq:apflux_hybrid}) & The ratio of the weighted average flux of $g$ and $r$ in the 3\farcs5 and 5\farcs0 apertures \\
\texttt{apratio\_gr}(5\farcs0, 7\farcs0) & (\ref{eq:feat_ap}), (\ref{eq:apflux_hybrid}) & The ratio of the weighted average flux of $g$ and $r$ in the 5\farcs0 and 7\farcs0 apertures \\
\texttt{apratio\_iz}(0\farcs5, 0\farcs75) & (\ref{eq:feat_ap}), (\ref{eq:apflux_hybrid}) & The ratio of the weighted average flux of $i$ and $z$ in the 0\farcs5 and 0\farcs75 apertures \\
\texttt{apratio\_iz}(0\farcs75, 1\farcs0) & (\ref{eq:feat_ap}), (\ref{eq:apflux_hybrid}) & The ratio of the weighted average flux of $i$ and $z$ in the 0\farcs75 and 1\farcs0 apertures \\
\texttt{apratio\_iz}(1\farcs0, 1\farcs5) & (\ref{eq:feat_ap}), (\ref{eq:apflux_hybrid}) & The ratio of the weighted average flux of $i$ and $z$ in the 1\farcs0 and 1\farcs5 apertures \\
\texttt{apratio\_iz}(1\farcs5, 2\farcs0) & (\ref{eq:feat_ap}), (\ref{eq:apflux_hybrid}) & The ratio of the weighted average flux of $i$ and $z$ in the 1\farcs5 and 2\farcs0 apertures \\
\texttt{apratio\_iz}(2\farcs0, 3\farcs5) & (\ref{eq:feat_ap}), (\ref{eq:apflux_hybrid}) & The ratio of the weighted average flux of $i$ and $z$ in the 2\farcs0 and 3\farcs5 apertures \\
\texttt{apratio\_iz}(3\farcs5, 5\farcs0) & (\ref{eq:feat_ap}), (\ref{eq:apflux_hybrid}) & The ratio of the weighted average flux of $i$ and $z$ in the 3\farcs5 and 5\farcs0 apertures \\
\texttt{apratio\_iz}(5\farcs0, 7\farcs0) & (\ref{eq:feat_ap}), (\ref{eq:apflux_hybrid}) & The ratio of the weighted average flux of $i$ and $z$ in the 5\farcs0 and 7\farcs0 apertures \\
\enddata
\tablecomments{All three models include \texttt{delta\_dchisq}(gal) and \texttt{mask}, and the major difference between the them is the type of aperture flux adopted in constructing \texttt{apratio}. The number of shared and unique features are indicated. The column Definition refers to the equations where the corresponding features are defined.}
\end{deluxetable*}

\section{Model Training}\label{sec:model}

\xgboost\ \citep{XGBoost_2016} is a robust and scalable machine learning system based on Gradient Boosted Decision Trees \citep{Friedman_2001}. It allows efficient handling of millions of sources, and naturally deals with missing values in the input with a sparsity-aware algorithm. It has been applied to tabulated data products of imaging surveys for multiple classification tasks \citep[e.g.,][]{Jin_2019, Nakoneczny_2021, Fu_2024, vonMarttens_2024, Zeraatgari_2024}, including star--galaxy separation \citep{Stoppa_2023}.

There are a couple of important hyperparameters in the \xgboost\ algorithm that we tune, and we determine the optimal hyperparameters with a broad grid search. The learning rate \texttt{eta} and the number of trees \texttt{n\_estimator} are coupled with each other. In general, lowering the learning rate improves model performance at the cost of computational resources as a larger number of trees is required. Once \texttt{eta} and all other hyperparameters are fixed, the optimal value for \texttt{n\_estimator} can be determined with early stopping, i.e., increasing the number of trees to the model until there is no performance improvement in the validation set. 

For the grid search, we explore the hyperparameters below, which are listed in order of descending importance for final model classifications:
\begin{itemize}
    \item \texttt{max\_depth} (the maximum depth of a tree which dominates) and \texttt{min\_child\_weight} (the minimum sum of instance weight needed in a child, below which further partition will be stopped): \{4, 6, 8, 10, 12, 14, 16, 18\} $\times$ \{4, 6, 8, 10, 12\}
    \item \texttt{gamma} (the minimum loss reduction required to make a further partition on a leaf node of the tree, which regularize the model complexity): \{0.1, 0.3, 0.5, 1.0, 2.0\}
    \item \texttt{subsample} (fraction of samples randomly drawn from the training instances) and \texttt{colsample\_bytree} (fraction of features randomly drawn in constructing each tree): \{0.4, 0.5, 0.6, 0.7, 0.8, 0.9\} $\times$ \{0.4, 0.5, 0.6, 0.7, 0.8, 0.9\}
\end{itemize}

In practice, we fix $\texttt{eta}=0.02$ and other parameters to a reasonable default value: $\texttt{max\_depth}=10$, $\texttt{min\_child\_weight}=6$, $\texttt{gamma}=0.1$, $\texttt{subsample}=0.8$, and $\texttt{colsample\_bytree}=0.8$ and estimate the smallest \texttt{n\_estimator} that does not show an appreciable drop in the FoM. With conduct a 5-fold cross validation (CV). In each of the 5 folds in the CV, we keep adding trees to models until the FoM on the validation set has no improvement with the addition of another 250 trees. Eventually, the median of the final tree numbers in the 5 folds is adopted as the optimal value. 

We tweak the other hyperparameters in sequence. We note that some parameters contribute to the model in a complementary way and need to be tuned jointly. \texttt{max\_depth} and \texttt{min\_child\_weight} both dominate the model complexity by controlling the vertical growth (depth) and horizontal growth (breadth) of individual trees, respectively. Deeper trees (higher \texttt{max\_depth}) might naturally need stricter \texttt{min\_child\_weight}, and shallower trees typically work better with more relaxed \texttt{min\_child\_weight}. Besides, \texttt{subsample} and \texttt{colsample\_bytree} both introduce the randomness to the model by drawing only a certain fraction of training samples and their features, respectively. Effectively, the fraction of the total data points used in each tree is the product of \texttt{subsample} and \texttt{colsample\_bytree}.
Considering both the relative importance of these hyperparameters and the way they are coupled with each other, we first explore a grid of \texttt{max\_depth} and \texttt{min\_child\_weight}. Once determined, they are fixed and we move to the subsequent parameter. Following the same rule, we then explore the regularization parameter \texttt{gamma}, and eventually \texttt{subsample} and \texttt{colsample\_bytree} jointly. 

The metric we aim to maximize is the figure of merit (FoM), defined as the true positive rate (TPR), i.e., the fraction of correctly identified stars, at a low fixed false positive rate (FPR), i.e., the fraction of misclassified galaxies. Following \cite{Miller_2017} and \citetalias{Tachibana_2018}, we set the FPR threshold to be 0.5\%. Our primary goal is to construct a complete catalog of host galaxies to identify extragalactic transients, and a maximized FoM ensures that our model can reject as many stars as possible while retaining most ($\sim$99.5\%) of the galaxies.

The optimization is conducted via 5-fold CV. Similar to \citetalias{Tachibana_2018}, we establish an inner and outer CV loop to ensure that the data used for hyperparameter tuning are not used for evaluating the model performance, and the sensitivity of our models to each of the hyperparameter. In the outer loop, we randomly split the training set into 5 partitions with the \texttt{StratifiedKFold} method in the \texttt{Python} library \texttt{scikit-learn} \citep{scikit-learn}, such that each split retains roughly the same fraction of stars and galaxies as the entire training set. For each partition in the outer CV loop, an inner 5-fold CV is applied to the remaining 80\% of the training data to determine the optimal hyperparameters. 
In each round of grid search, we will make predictions on the sources withheld in the outer loop with the optimal model from the inner loop to produce the scores (probability of being a point source) for every target in the training set. The outputs in the last round are used to evaluate the model performance on the training set. 
The median value of each parameter over the 5 folds is adopted as the optimal value, which is fixed in subsequent rounds. 
We note that the FoM evaluated on the sources withheld in the outer loop does not increase significantly once \texttt{n\_estimator} is determined before the first round of the grid search.
This indicates that the FoM is not highly sensitive to the choice of the hyperparameters controlling the depth of the model, so we refrain from exploring a finer grid.

The weight in the hybrid model, $\alpha_\mathrm{hyb}$, is determined separately in an iterative way. To begin with, we presume the two sub-models in the hybrid models both adopt the default hyperparameters, with which we predict the scores for both sub-models using the same 5-fold CV. Then we figure out the $\alpha_\mathrm{hyb}$ to give rise to the highest mean FoM over the 5 validation sets by searching a grid from 0 to 1 with a step size of 0.01. Fixing $\alpha_\mathrm{hyb}$, we then search for the optimal \xgboost\ parameters using the same method, with which we update the value of $\alpha_\mathrm{hyb}$. In practice, $\alpha_\mathrm{hyb}$ converges to 0.33 after the second round, which yields an FoM $= 0.707\pm0.011$. Yet for all $\alpha_\mathrm{hyb}\simeq0.2$--0.5, the corresponding FoM does not significantly drop, i.e., remains within the standard deviation of the optimal value. {To assess the robustness of our chosen $\alpha_\mathrm{hyb}$, we determine the optimal $\alpha_\mathrm{hyb}$ value in the DESI and SDSS test sets, fixing all other hyperparameters and using the same procedure as described above. Both test sets cover a much wider region of the sky than the training set. For DESI (SDSS), the highest FoM is 0.9792 (0.9718) at $\alpha_\mathrm{hyb}= 0.51$ (0.57), which represents a minor ($\sim$0.1\%) improvement relative to the FoM $=$ 0.9778 (0.9704) at $\alpha_\mathrm{hyb}= 0.33$. This demonstrates that the model performance is not strongly dependent on the choice of $\alpha_\mathrm{hyb}$, given that the FoM does not change substantially when $\alpha_\mathrm{hyb}$ changes from 0.33 to $\sim$0.5. For the final model we adopt $\alpha_\mathrm{hyb} = 0.33$ to make the final catalog classifications.

}





        


The optimal hyperparameters are listed in Table~\ref{tab:hp}, and we note again that the model performance is not highly sensitive to the parameter choices.

\begin{deluxetable}{cccc}\label{tab:hp}
\tablehead{
    \colhead{Parameter} & \colhead{White} & \colhead{$grz$} & \colhead{Hybrid}
}
\tablecaption{Hyperparameters adopted in each \xgboost\ model.}
\startdata
\texttt{eta} & 0.02 & 0.02 & 0.02 \\
\texttt{n\_estimator} & 850 & 1300 & 800 \\
\texttt{max\_depth} & 12 & 18 & 12 \\
\texttt{min\_child\_weight} & 6 & 10 & 6 \\
\texttt{gamma} & 1.0 & 0.1 & 0.5 \\
\texttt{subsample} & 0.5 & 0.8 & 0.4 \\
\texttt{colsample\_bytree} & 0.8 & 0.4 & 0.9 \\
$\alpha_\mathrm{hyb}$ & \dots & \dots & 0.33
\enddata
\end{deluxetable}

\section{Model Performance}\label{sec:results}
\subsection{Performance on the COSMOS Field}\label{sec:res_cosmos}
\begin{figure*}
    \centering
    \includegraphics[width=\linewidth]{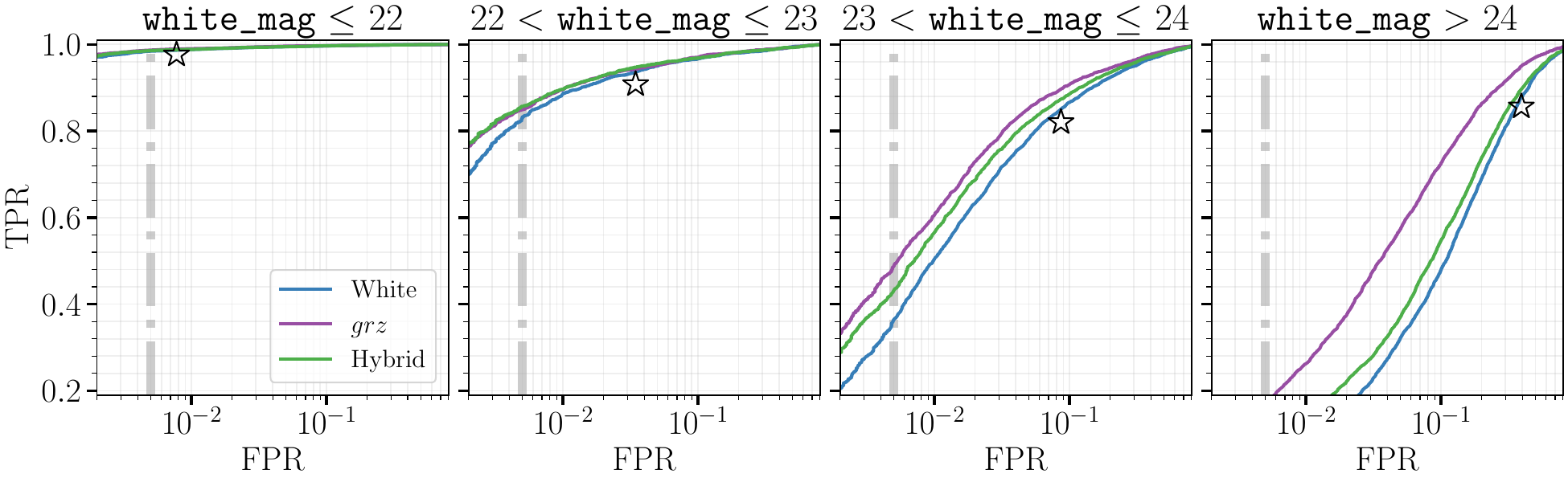}
    \caption{ROC curves of the three \xgboost\ models (White, $grz$, and Hybrid) as evaluated in the training set by CVs in different magnitude bins. Models including aperture fluxes from individual filters ($grz$ and Hybrid) outperforms the White model which only includes the average aperture flux of all filters, especially for faint sources. The thick lines show the performance of each model on the entire training set.
    The empty black asterisk corresponds to the TPR and FPR if the morphological types provided by LS are adopted. The dashed vertical line correspond to an FPR of 0.5\%, at which we evaluate the FoM.}
    \label{fig:roc_mag}
\end{figure*}

\begin{figure}
    \centering
    \includegraphics[width=\linewidth]{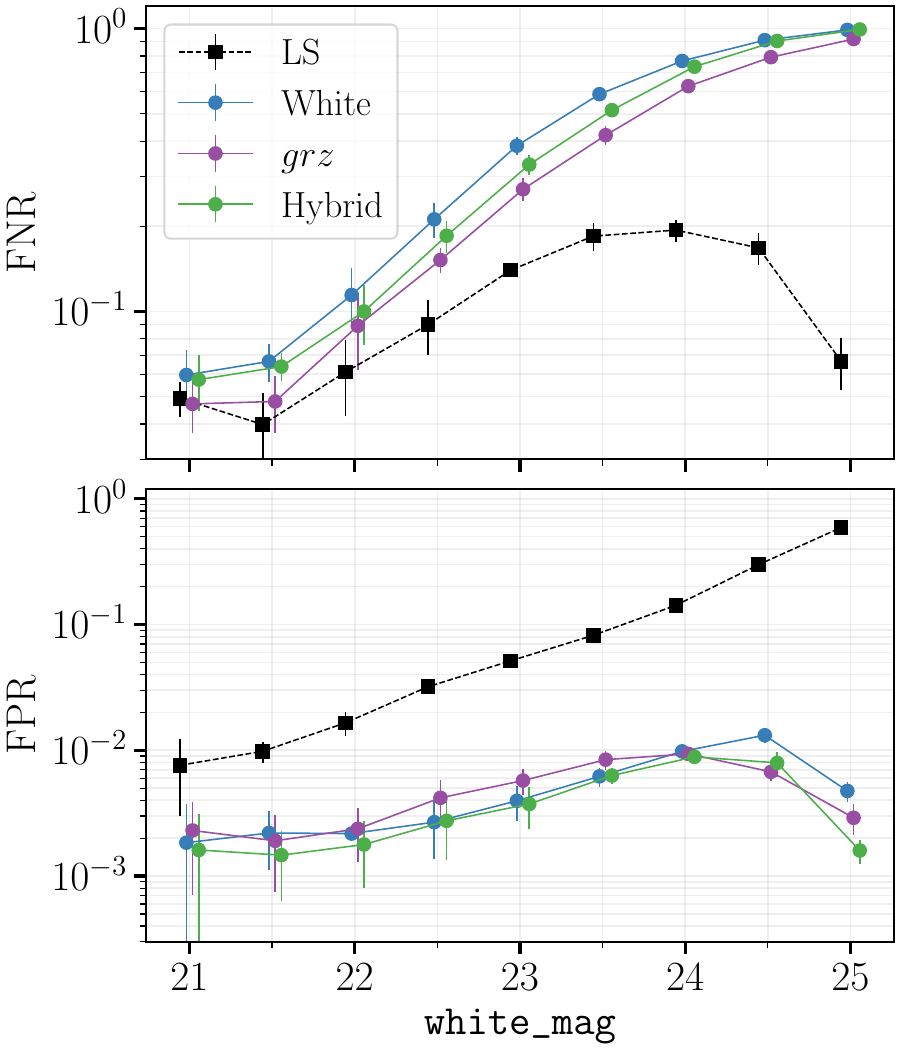}
    \caption{The FPR and FNR (the fraction of misclassified stars and galaxies) as a function of \texttt{white\_mag} of the LS model (dashed black lines) and our \xgboost\ models (solid blue, purple, and green lines). The bin widths are 0.5\,mag. Each marker corresponds to the FPR/FNR evaluated on the entire dataset using CV, while the error bar indicates the sample standard deviation over the 5 CV folds. Note that the markers corresponding to the same bin centers are slightly offset for clarity.}
    \label{fig:FPFN_mag}
\end{figure}

The COSMOS field is one of the most thoroughly observed areas in the LS footprint. The metrics evaluated on the HST training set thus reflect the model performance under the ideal circumstances: (i) the stacked images yield deep detection limits; (ii) almost all targets have been visited in all of $griz$ filters. 

In Table~\ref{tab:hst} we list the FoM, accuracy,\footnote{The fraction of correctly classified objects including TPs and TNs $$\mathrm{Acc = \frac{TP + TN}{P + N}}.$$} and the area under the ROC\footnote{The TPR-FPR pairs at all classification thresholds.} (receiver operating characteristic) curve (ROC AUC) of the 3 models we consider, evaluated on the training set using 5-fold CV. For all \xgboost\ models, we evaluate the accuracy at a classification threshold of 0.5, unless otherwise specified. The uncertainties represent the sample standard deviation over the 5 folds. We find that the $grz$ model outperforms other models in all three metrics, whereas the White model always performs the worst among the three. The models all outperform the random-forest model in \citetalias{Tachibana_2018} when validated on the training set. While the CV results in the \citetalias{Tachibana_2018} training set seem to be comparable (e.g., FoM = $0.707\pm0.036$ evaluated at the same threshold FPR = 0.5\%), their dataset contains only the brightest sources of the training set in this work. For an apple-to-apple comparison, in each of our 5 validation sets, we pick out sources that are also included in the \citetalias{Tachibana_2018} training set, with which evaluate the FoM and its uncertainty. The White, $grz$, and Hybrid models yield an FoM (the mean and sample standard deviation for the 5 validation sets) of $0.961\pm0.007$, $0.966\pm0.008$, and $0.966\pm0.006$, respectively, all of which show a gigantic boost relative to the performance of the \citetalias{Tachibana_2018} model.\footnote{In practice we only use 46,383 out of 47,093 sources in \citetalias{Tachibana_2018}, which have an LS \dr{10} counterpart within 1\arcsec.}

In Figure~\ref{fig:roc_mag} we show the corresponding ROC curves in different magnitude bins. It is worth noting that compared to separating stars/galaxies using the morphological types in LS (referred to as the LS model hereafter, which is defined as a binary classifier treating targets with the PSF label as stars, and all other labels as galaxies), our \xgboost\ models always recover a higher fraction of stars when missing the same amount of galaxies. This is expected because the morphological labels are calculated based on the \texttt{dchisq} scores only, whereas our \xgboost\ models contain additional information from the aperture photometry.
All three \xgboost\ models show a near-perfect ability of classifying resolved--unresolved sources for targets brighter than $\texttt{white\_mag}=22$\,mag, with their ROC curves being almost identical. For fainter targets closer to the detection limit of LS, the $grz$ model outperforms both the Hybrid and the White model.

The difference in model performance on faint objects ($\texttt{white\_mag}\ge21$\,mag) are further addressed in Figure~\ref{fig:FPFN_mag}. We show the fraction of misclassified stars and galaxies of the \xgboost\ models, i.e., the false negative rate (FNR) and the FPR at a classification threshold of 0.5, compared to the LS model. The \xgboost\ models correctly identify $\gtrsim$99\% of galaxies across all magnitude bins at the expense of missing the majority of faint stars ($\texttt{white\_mag}\ge24$\,mag). In contrast, the LS model misclassifies more than 20\% of the galaxies, while successfully recovering $\gtrsim$80\% of the stars. The different approach of each model explains this discrepancy in performance. The \xgboost\ models are trained by minimizing the loss on the training set as a whole, whereas LS morphological types are determined via a hypothesis-test for each source. The null hypothesis is that each source is an unresolved PSF and the LS morphological model only rejects the hypothesis when extended profiles significantly improve the goodness of fit. As shown in Figure~\ref{fig:bias} nature produces far more faint galaxies than stars so the \xgboost\ models will tend towards the majority class. Meanwhile, for the faintest targets improvements in the goodness of fit (\texttt{dchisq}) are limited by the S/N, making it progressively more difficult to reject the PSF null hypothesis. Thus, we expect our \xgboost\ classifiers to show a much higher overall accuracy compared to the LS model, especially for faint targets (see Figure~\ref{fig:acc}).

While the \xgboost\ models are generally not good at identifying faint stars, Figures~\ref{fig:roc_mag} and~\ref{fig:FPFN_mag} suggest that knowledge from individual filters can boost model performance. Between 24--25\,mag, the $grz$ model consistently identifies $>$10\% more stars than the White model at the same FPR.  {In Appendix~\ref{sec:feature_importance} we list the feature importance, defined as the average gain in accuracy when a feature is used in a split, of all the features in the three models. In the $grz$ model, the top 5 features that bring the greatest improvement are:} \texttt{mask} (see Equation~\ref{eq:feat_mask}), \texttt{delta\_dchisq}(EXP), \texttt{delta\_dchisq}(REX), \texttt{delta\_dchisq}(DEV) (improvement in the fit by substituting the PSF profile with EXP, REX, and DEV profiles, respectively), and $\texttt{Feat[ap\_grz]}(z, 5)$ (the flux ratios in $z$ between the 2\farcs0 and 3\farcs5 apertures). Interestingly, the feature importance of the flux ratios in $g$ and $r$ filters are a few times lower. This is probably due to the dramatic color differences between faint stars (predominantly M-dwarfs) and galaxies as illustrated in Figure~\ref{fig:bias}. Since the Galactic reddening in the COSMOS field is minimal and approximately constant, the colors of stars and galaxies approximate their intrinsic distributions. The extremely red $g-z$ and $r-z$ colors of M dwarfs means that $z$ is often the only filter with sufficient S/N to identifying them. Merging fluxes from all filters, as is done in the White model, introduces some level of noise from other filters with lower S/N.

In conclusion, when photometry is available in all bands, including features based on the aperture flux in individual filters improves model performance. In this ideal situation, the $grz$ model (three filters) is always optimal, and the Hybrid model (effectively two broad-band filters) always outperforms the White model (all filters merged).

\begin{deluxetable*}{ccccc}[h]
\tablehead{
    \colhead{Metric} & \colhead{LS} & \colhead{White} & \colhead{$grz$} & \colhead{Hybrid}
}
\tablecaption{HST Training Set Metrics.\label{tab:hst}}
\startdata
    FoM      & $\cdots$ & 0.668 $\pm$ 0.012 & {\bf 0.740} $\pm$ 0.010 & 0.707 $\pm$ 0.011 \\
    Accuracy & 0.797 $\pm$ 0.002 & 0.967 $\pm$ 0.001 & {\bf 0.973} $\pm$ 0.001 & 0.970 $\pm$ 0.001 \\
    ROC AUC  & $\cdots$ & 0.962 $\pm$ 0.002 & {\bf 0.976} $\pm$ 0.001 & 0.967 $\pm$ 0.002 \\
\enddata
\tablecomments{Uncertainties represent the sample standard deviation for the 5 folds used in CV. Values in bold indicate the model with the best performance for each metric.}
\end{deluxetable*}

\subsection{Performance on the Test Sets}\label{sec:test_performance}

\begin{figure*}
    \centering
    \includegraphics[width=0.83\linewidth]{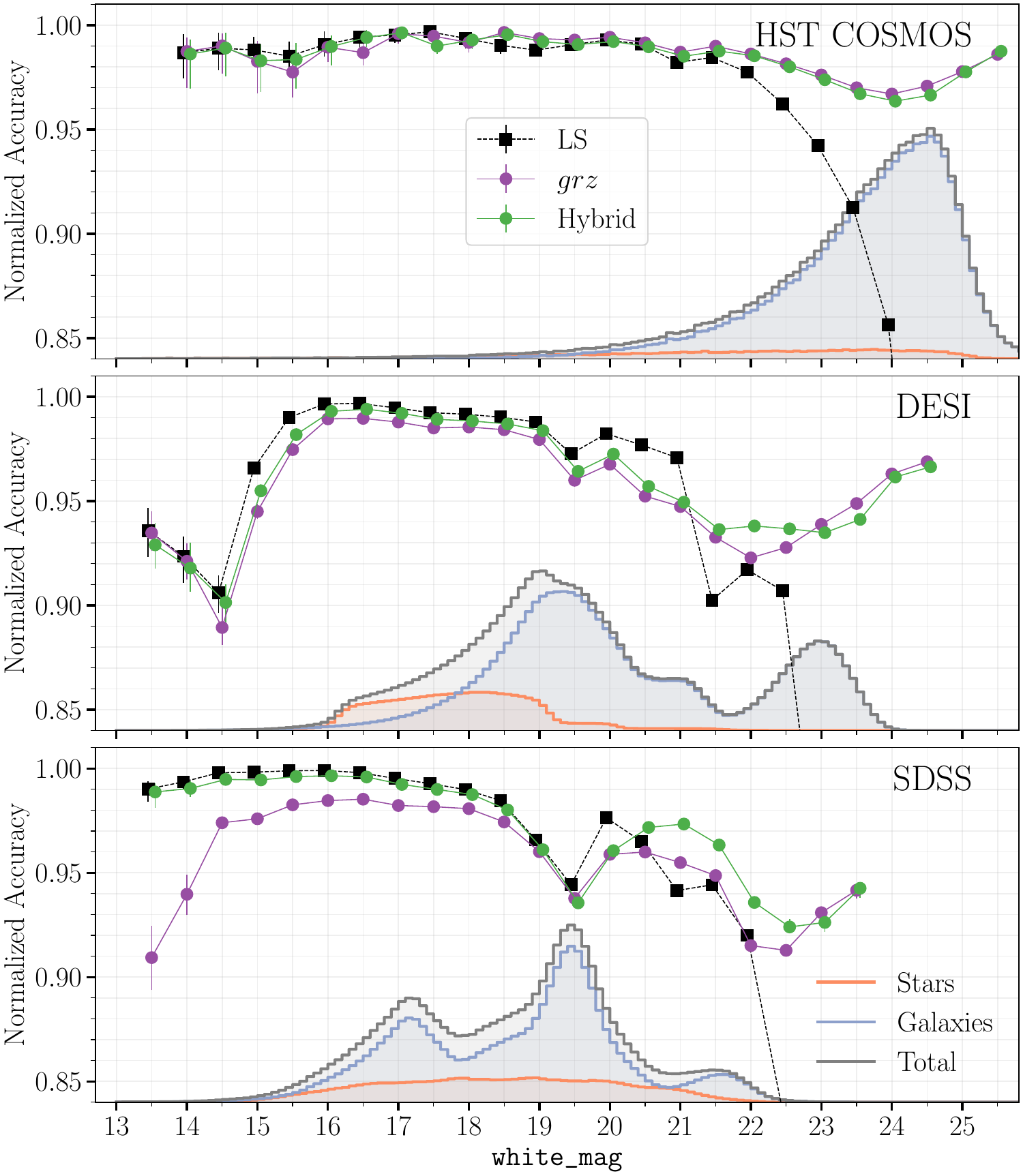}
    \caption{Accuracy as a function of the \texttt{white\_mag} evaluated in the HST training set (using CV) and two independent test sets (DESI and SDSS). The bin widths are 0.5\,mag, and in each bin, the number of stars and galaxies have been normalized to match their relative ratio in the full HST$\times$LS dataset (see Figure~\ref{fig:bias}) 
    Uncertainties of the normalized accuracy are defined in the text. The original distribution of star and galaxy counts in each dataset is also displayed. The markers corresponding to the same bin centers are slightly offset for clarity.}
    \label{fig:acc}
\end{figure*}

To test the model performance in non-ideal conditions (e.g., incomplete filter coverage, poor seeing, reduced depth, and significant Galactic extinction), we evaluate the same metrics (FoM, accuracy, ROC AUC) on the DESI and SDSS test sets, both covering a much wider field than the HST training set. The results are displayed in Table~\ref{tab:test}. In sharp contrast to the CV results in the COSMOS field, the Hybrid model clearly outperforms the other \xgboost\ models in all metrics. Uncertainties represent the sample standard deviation for 100 bootstrap samples (with replacement) of each dataset. The FoM of the $grz$ is still comparable to that of the White model on the DESI dataset, but for SDSS sources, the $grz$ model has the worst performance, suggesting that it is not robust against non-ideal data quality.

In Figure~\ref{fig:acc}, we show the accuracy of the LS model as well as the $grz$ (best FoM on the HST training set) and Hybrid models (best FoM on the DESI and SDSS test sets) on both the HST training set and the DESI and SDSS test sets. Note that the spectroscopic test sets are both composed of multiple dedicated surveys on certain types of sources (e.g., QSOs, ELGs, LRGs). The spectroscopic selection function for both SDSS and DESI results in a sample that is very different from the overall LS dataset. For example, galaxies are over-represented in the DESI catalog on both the bright ($\texttt{white\_mag}\lesssim16$\,mag) and faint ($\texttt{white\_mag}\gtrsim19$\,mag) ends.
We overlay the distributions of star and galaxy counts in each dataset on Figure~\ref{fig:acc} which clearly shows the relative counts of stars to galaxies. By definition, the overall accuracy depends on the TPR, TNR, and the stellar fraction $p_s$,
\begin{equation}\label{eq:acc}
    \mathrm{Acc} = p_s\times\mathrm{TPR} + (1-p_s)\times\mathrm{TNR}.
\end{equation}
To estimate the accuracy with a biased star--galaxy ratio, we correct for the number count bias via a Monte Carlo approach. In a magnitude bin of $N$ objects from the HST$\times$LS dataset (whose source distributions are assumed as a good representative of the entire LS catalog), if the star counts $N_s$ is Binomial-distributed with a probability of $p_s$, the conjugate prior on $p_s$ is a Beta distribution, 
\begin{equation*}
    p_s\sim\mathcal{B}(N_s+1, N-N_s+1).
\end{equation*}
Similarly, we assume the TPR and TNR in the same magnitude bin of the biased dataset, where there are $n_s$ stars ($=$$\mathrm{TP+FN}$) and $n-n_s$ galaxies ($=$$\mathrm{TN+FP}$), are also Beta-distributed,
\begin{equation*}
    \begin{aligned}
        \mathrm{TPR}&\sim\mathcal{B}(\mathrm{TP}+1, \mathrm{FN}+1),\\
        \mathrm{TNR}&\sim\mathcal{B}(\mathrm{TN}+1, \mathrm{FP}+1).
    \end{aligned}
\end{equation*}
Finally, we draw 100 independent samples of $p_s$, TPR, and TNR to produce a sample of the accuracy using Equation~(\ref{eq:acc}), and what we present in Figure~\ref{fig:acc} are the median and 68\% credible interval of the accuracy in these samples.

Generally speaking, all three models effectively separate bright stars from galaxies, but for sources fainter than $\sim$22\,mag, the accuracy of the LS model dramatically drops due to its poor ability of identifying galaxies, while the accuracy of the \xgboost\ models remains above 90\%. Compared to the smooth accuracy curves evaluated on the HST training set via CV, the corresponding curves on the DESI and SDSS datasets feature a couple of dips ($\lesssim$15\,mag, around $\sim$19.5\,mag, and around $\sim$22.5\,mag). 

Stars $\lesssim$15\,mag can easily saturate LS images. The top-hat profiles of these saturated point sources sometimes fit better to extended morphological models, leading to misclassification. To identify bright stars and mask them properly, LS uses Gaia as a reference catalog. In LS \dr{10}, at the coordinates of each bright source ($\le$18\,mag) in Gaia early data release 3 (E\dr{3}), an LS source will be modeled and extracted with only a PSF profile. These sources will be flagged as a \texttt{GAIA\_POINTSOURCE} in their \texttt{FITBITS} bitmask. In our \xgboost\ classifiers, their \texttt{mask} will be set to 2 (Equation~\ref{eq:feat_mask}), which supports the classification despite the saturation. In the northern footprint (declination $>32.375\degr$) where the LS \dr{9} results are adopted, the selection of Gaia point sources was based on Gaia \dr{2}, and an additional astrometric quality cut $\texttt{astrometric\_excess\_noise}<\sqrt{10}$ was applied. A handful of bright LS stars are associated with Gaia sources which failed to pass the astrometric quality cut, mostly in the DESI footprint, and are not flagged properly as a result. They become the dominant FNs in the DESI test set when $\texttt{white\_mag} \le 15$\,mag, causing a dramatic drop in the classification accuracy (Figure~\ref{fig:acc}). In Section~\ref{sec:pipeline} we will show how we tweak the scores for these bright stars identified in Gaia \dr{3} to mitigate this issue. Additionally, our real-time data processing pipeline for LS4, which only makes use of LS sources in the southern footprint, will not be impacted. 

The kink at $\sim$19.5\,mag in the SDSS dataset has been discovered and well studied in \cite{Miller_2017} and \citetalias{Tachibana_2018}, which is due to a disproportionately large number of blended K/M-type stars targeted as LRG candidates in SDSS. These sources are typically unresolved in seeing-limited, ground-based optical images. Similarly in the DESI catalog, blended K/M-type stars are over-represented both as BGS \citep[bright galaxy survey;][]{DESI_BGS_2023} and LRG candidates, corresponding to the valleys around 19.5\,mag and 21.5\,mag, respectively.

We also find that our model has a relative high FNR on the bluest stars ($g-z<0$\,mag), such as white dwarfs (WDs). On the LS imaging these blue sources are mostly isolated objects, so the misclassification is not due to blending. Instead, their low S/N on the red probably hinders the classification, because features containing aperture flux from red filters are vital to the \xgboost\ models as demonstrated in Section~\ref{sec:res_cosmos}. In fact, unlike blended stars whose score distribution peaks sharply at around 0, scores of these misclassified faint blue dots are more uniformly distributed from 0 to 0.5, suggesting a high level of classification ambiguity. 

When $\texttt{white\_mag}\gtrsim 23.5$\,mag, the stellar fraction $p_s$ drops below 5\% (Figure~\ref{fig:bias}). In this magnitude range, our \xgboost\ classifiers are insensitive to stars and tend to classify everything as a galaxy, so the normalized accuracy is asymptotically the galaxy fraction in the full HST$\times$LS dataset, $1-p_s$. The trend of increasing accuracy for fainter object in all three datasets is a result of the decreasing $p_s$.

In the entire LS dataset where the source distributions are not subject to the selection biases in spectroscopic surveys, either blended K/M-type stars or faint WDs should not make up any substantial fraction. We thus argue that the performance of the \xgboost\ models should exceed their current appearance on the test sets. We also conclude that for objects fainter than $\sim$22\,mag, the \xgboost\ models outperform the LS model. While the $grz$ model slightly outperforms the Hybrid model on the faintest objects $\gtrsim$23\,mag on the test set, its generalization ability on datasets with non-ideal quality is not comparable to the Hybrid model, as discussed below.

\begin{deluxetable*}{cccccc}[h]
\tablehead{
    \colhead{Test Set (Size)} & \colhead{Metric} & \colhead{LS} & \colhead{White} & \colhead{$grz$} & \colhead{Hybrid}
}
\tablecaption{DESI and SDSS Test Sets Metrics.\label{tab:test}}
\startdata
       &FoM      &$\cdots$& 0.9712 $\pm$ 0.0002 & 0.9717 $\pm$ 0.0001 & {\bf 0.9778} $\pm$ 0.0001 \\
 DESI (15,354,071) &Accuracy &0.9451 $\pm$ 0.0001& 0.9898 $\pm$ 0.0001 & 0.9905 $\pm$ 0.0001 & {\bf 0.9919} $\pm$ 0.0001 \\
       &ROC AUC  &$\cdots$& 0.99750 $\pm$ 0.00002 & 0.99768 $\pm$ 0.00002 & {\bf 0.99831} $\pm$ 0.00002 \\
       \hline
       &FoM      &$\cdots$& 0.9551 $\pm$ 0.0004 & 0.9411 $\pm$ 0.0007 & {\bf 0.9704} $\pm$ 0.0004\\
 SDSS (3,357,255)  &Accuracy & 0.9842 $\pm$ 0.0001 & 0.9854 $\pm$ 0.0001 & 0.9818 $\pm$ 0.0001 & {\bf 0.9858} $\pm$ 0.0001 \\
       &ROC AUC  &$\cdots$& 0.99704 $\pm$ 0.00005 & 0.99643 $\pm$ 0.00005 & {\bf 0.99773} $\pm$ 0.00004
\enddata
\tablecomments{Uncertainties represent the sample standard deviation for 100 bootstrap samples of each dataset. Values in bold indicate the model with the best performance for each metric.}
\end{deluxetable*}

\subsubsection{Missing Filters}
Given the importance of aperture flux ratios in the red filters, we test the models on a subset of sources with both the $i$ and $z$ filters missing and list the results in Table~\ref{tab:test_no_iz}. In the absence of $i$ and $z$ filters, the performance of all models degrades, especially for the $grz$ model. The Hybrid model has similar performance as the White model in terms of FoM and accuracy, while the $grz$ model shows $\sim$20\% worse accuracy. By definition, when $z$-band data are missing, values for all $\texttt{Feat[ap\_grz]}(z, k)$ are set to NaN. \xgboost\ handles sparsity in the input data by learning a default direction in each tree node from the training set. Then when a value is missing, the instance is classified in the default direction \citep{XGBoost_2016}. In the HST training set, only faint objects close to the $z$-band detection limit, which are predominantly galaxies, will be missing $\texttt{Feat[ap\_grz]}(z, k)$. On the test sets, the $grz$ model would therefore assign a systematically lower score whenever $\texttt{Feat[ap\_grz]}(z, k)=\mathrm{NaN}$, including bright stars that are simply not covered in $z$. The overall performance of the $grz$ model thus heavily depends on the fraction of objects missing $z$-band data. On the entire LS dataset, $\sim$$2\times10^7$ sources ($\sim$7\%) are not covered by $z$ filter, among which the $grz$ model will be highly biased. 

\begin{figure*}[h]
    \centering
    \includegraphics[width=0.83\linewidth]{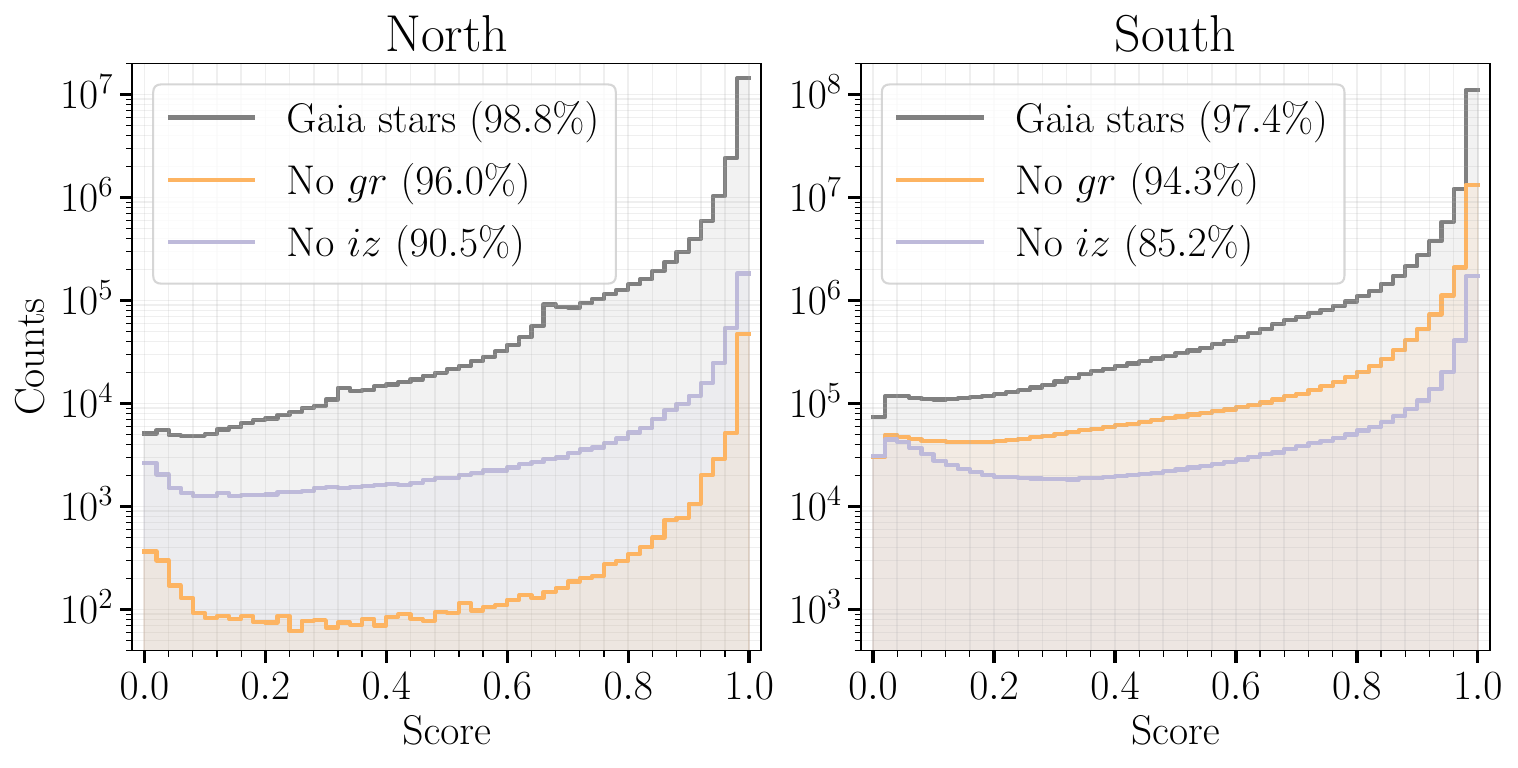}
    \caption{Score distribution of the Hybrid model for Gaia stars in the northern (left) and southern (right) LS footprint. The grey histograms correspond to the entire dataset, whereas the orange (purple) histograms present the subset of sources missing photometry in both $g$ and $r$ ($i$ and $z$). The TPR in each subset is indicated in the legend.}
    \label{fig:score_hist}
\end{figure*}

By construction, the Hybrid model has exactly the same features as the White model when either $gr$ or $iz$ filters are both missing. In Figure~\ref{fig:score_hist} we display the score distribution of the Hybrid model for Gaia stars, which has all-sky coverage up to $\sim$21\,mag. The fraction of stars with incomplete filter coverage should thus be representative of the entire LS dataset. The results again suggest a near-perfect classification accuracy for bright stars, even if a non-negligible fraction of stars are subject to missing filters. Among the Gaia stars with extremely low scores ($\le$0.1), almost half of them do not have either $gr$ or $iz$ data. Nevertheless, the decrease in TPR is mild. We therefore conclude that the Hybrid model is robust against missing filters. In the published catalog, we adopt scores from the Hybrid model.

\begin{deluxetable*}{cccccc}[h]
\tablehead{
    \colhead{Test Set (Size)} & \colhead{Metric} & \colhead{LS} & \colhead{White} & \colhead{$grz$} & \colhead{Hybrid}
}
\tablecaption{DESI and SDSS Test Sets Metrics for Sources missing both $iz$ filters.\label{tab:test_no_iz}}
\startdata
             &FoM      &$\cdots$& {\bf 0.632} $\pm$ 0.019 & 0.094 $\pm$ 0.010 & 0.603 $\pm$ 0.018  \\
 DESI (6,486)  &Accuracy &0.491 $\pm$ 0.008& 0.907 $\pm$ 0.005 & 0.736 $\pm$ 0.007 & {\bf 0.909} $\pm$ 0.005 \\
             &ROC AUC  &$\cdots$& 0.927 $\pm$ 0.006 & 0.817 $\pm$ 0.007 & {\bf 0.932} $\pm$ 0.005 \\
             \hline
             &FoM      &$\cdots$& 0.812 $\pm$ 0.028 & 0.380 $\pm$ 0.074 &  {\bf 0.820} $\pm$ 0.026\\
 SDSS (4,750) &Accuracy & 0.921 $\pm$ 0.005 & {\bf 0.937} $\pm$ 0.005 & 0.704 $\pm$ 0.011 & 0.934 $\pm$ 0.005 \\
             &ROC AUC  &$\cdots$& {\bf 0.984} $\pm$ 0.002 & 0.963 $\pm$ 0.003 & 0.983 $\pm$ 0.002
\enddata
\tablecomments{Uncertainties represent the sample standard deviation for 100 bootstrap samples of each subset missing $iz$ filters. Values in bold indicate the model with the best performance for each metric.}
\end{deluxetable*}

\subsubsection{Shallow Fields {and Seeing Variation}}

\begin{figure*}
    \centering
    \includegraphics[width=0.83\linewidth]{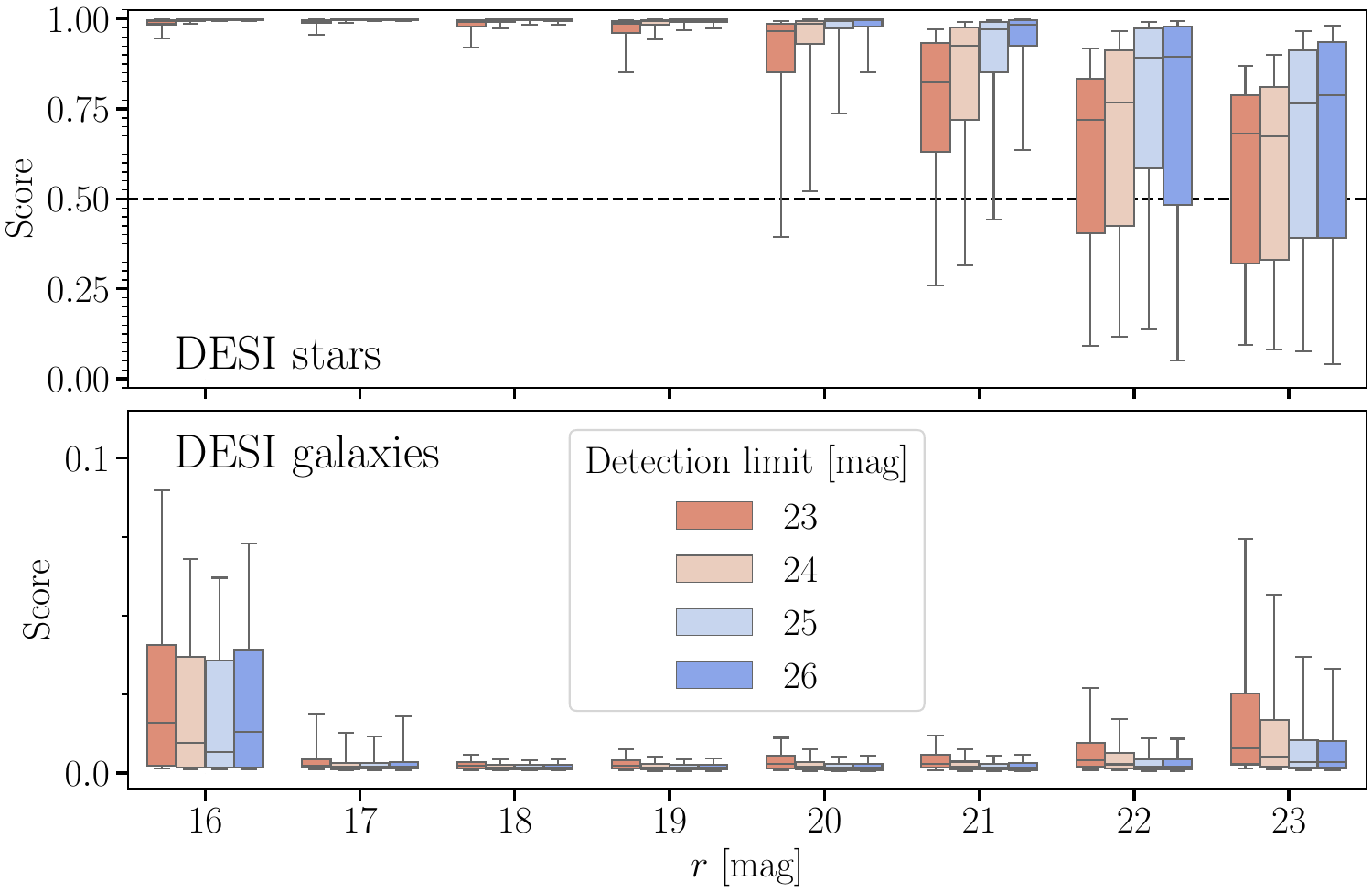}
    \caption{Box plots of model scores showing how the Hybrid model downgrades for fainter sources (16--23\,mag; from left to right) in different fields with a variety of 5-$\sigma$ detection limits (23--26\,mag; from red to blue colors) in LS. Each box presents the distribution of scores for a group of DESI stars (upper panel) and galaxies (lower panel) within a bin of certain magnitude and detection limit in $r$. All the bin sizes are 1\,mag$\times$1\,mag. All bins contain more than 500 objects. The horizontal line inside the box marks the median. The upper and lower edges of each box correspond to the 25th and 75th percentiles of the scores, and the whiskers extend to the 10th and 90th percentiles. We note that the lower panel is zoomed in, and the minor ticks in both panels correspond to a same score interval of 0.025. Many stars near the detection limit are misclassified with a score below 0.5, the dashed horizontal line.}
    \label{fig:maglim}
\end{figure*}

{
\begin{figure*}
    \centering
    \includegraphics[width=0.83\linewidth]{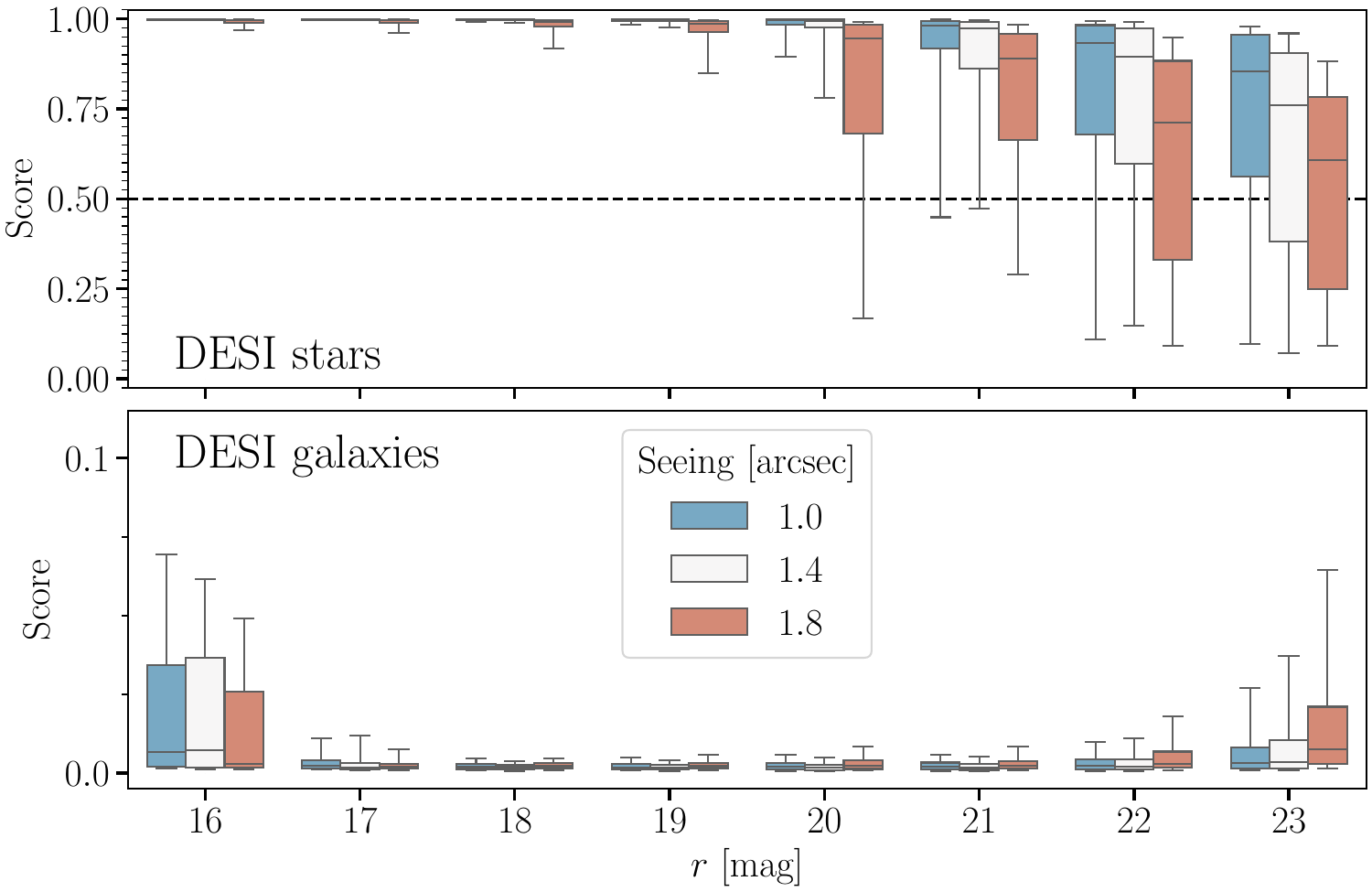}
    \caption{Similar as Figure~\ref{fig:maglim}, but displaying how the Hybrid model performs in different fields with a variety of the seeing (measured as the PSF size in $r$) in LS. To alleviate the correlation between the seeing and the limiting magnitude, in this plot we only show sources in the fields where the 5-$\sigma$ detection limits in $r$ are between 24.5 and 25.5\,mag.}
    \label{fig:seeing}
\end{figure*}
}

The CV results in the COSMOS field have revealed that faint stars near the detection limit are systematically misclassified as galaxies by the \xgboost\ models. This is acceptable since the marginally detected targets in the deepest LS fields (5-$\sigma$ detection limits in $r\gtrsim25$\,mag), including the COSMOS field, should be predominantly galaxies, with a stellar fraction no more than a few percent (see Figure~\ref{fig:bias}). However, the survey depth is highly non-uniform across the entire footprint, due to (i) variable observing conditions (e.g., the telescope aperture, seeing, airmass), and (ii) a different total number of visits for each location within the survey footprint. It is worth testing whether low-S/N targets in shallower fields, among which the stellar fraction is not negligible, suffer the same systematics.

Additionally, as the LS were conducted by telescopes at multiple sites, the seeings also vary significantly across the sky, affecting the PSF, i.e., the flux distribution across aperture sizes. As a result, our model, trained specifically on sources with a nearly constant seeing ~1'', is likely suboptimal in regions where seeings are much worse. A bad seeing also aggravates the stellar blending.

In Figure~\ref{fig:maglim}, we present box plots showing the distribution of scores among DESI stars and galaxies of different brightness in fields with different detection limits. The score is generated by the Hybrid model. From left to right, we show the distribution of model scores for DESI test set sources in 4 bins with LS 5$\sigma$ detection limits ranging from $r = 22.5$ to $26.5$\,mag. These four bins, each 1\,mag wide, cover $\sim$99.9\% of the targets in the DESI test set. 

An obvious trend is that more objects (both stars and galaxies) are assigned a score closer to 0.5 in fainter bins, indicating increasing ambiguity in classification. {The only exception is that many bright galaxies ($r\le16$\,mag) have scores between 0.05 and 0.1, leading to a larger spread of the galaxy scores in the first magnitude bin of Figure~\ref{fig:maglim}.} Most of these galaxies are {either} highly extended {or has a bright unresolved nucleus, such that} \texttt{Tractor} splits the galaxy into {multiple sources}. The scores reported here are for only the nuclei leading to slightly higher scores than a typical bright galaxy. {Despite the larger average scores, we do not see a drop in the FPR in classifying the brightest galaxies when adopting a classification threshold of 0.5.}

In each brightness bin, the score of all DESI galaxies and stars depends largely on the detection limit. In shallower fields, as the S/N is generally lower, both stars and galaxies receive a score closer to 0.5. This indicates that while the star--galaxy separation becomes trickier, decreasing the S/N does not bias our model towards lower scores, and low S/N galaxies are clearly receiving higher scores. 

{Figure~\ref{fig:seeing} displays how the seeing impacts the score distribution. To eliminate the effect of the imaging depth, we only include sources in the bricks with a 5$\sigma$ detection limit of 23.5 -- 24.5\,mag in $r$. As expected, the model performance decreases as the seeing degrades for both stars and galaxies. This occurs due to a combination of suboptimal model parameters and more severe blending.}

Consequently, we conclude there is no evidence for {significant} systematics that will bias the fidelity of the Hybrid model as the detection limit {or seeing} varies{, although the model parameters are likely suboptimal in the fields with seeings $\gg$ 1\arcsec}.

\subsubsection{Stellar Density}\label{sec:density}
\begin{figure*}
    \centering
    \includegraphics[width=0.83\linewidth]{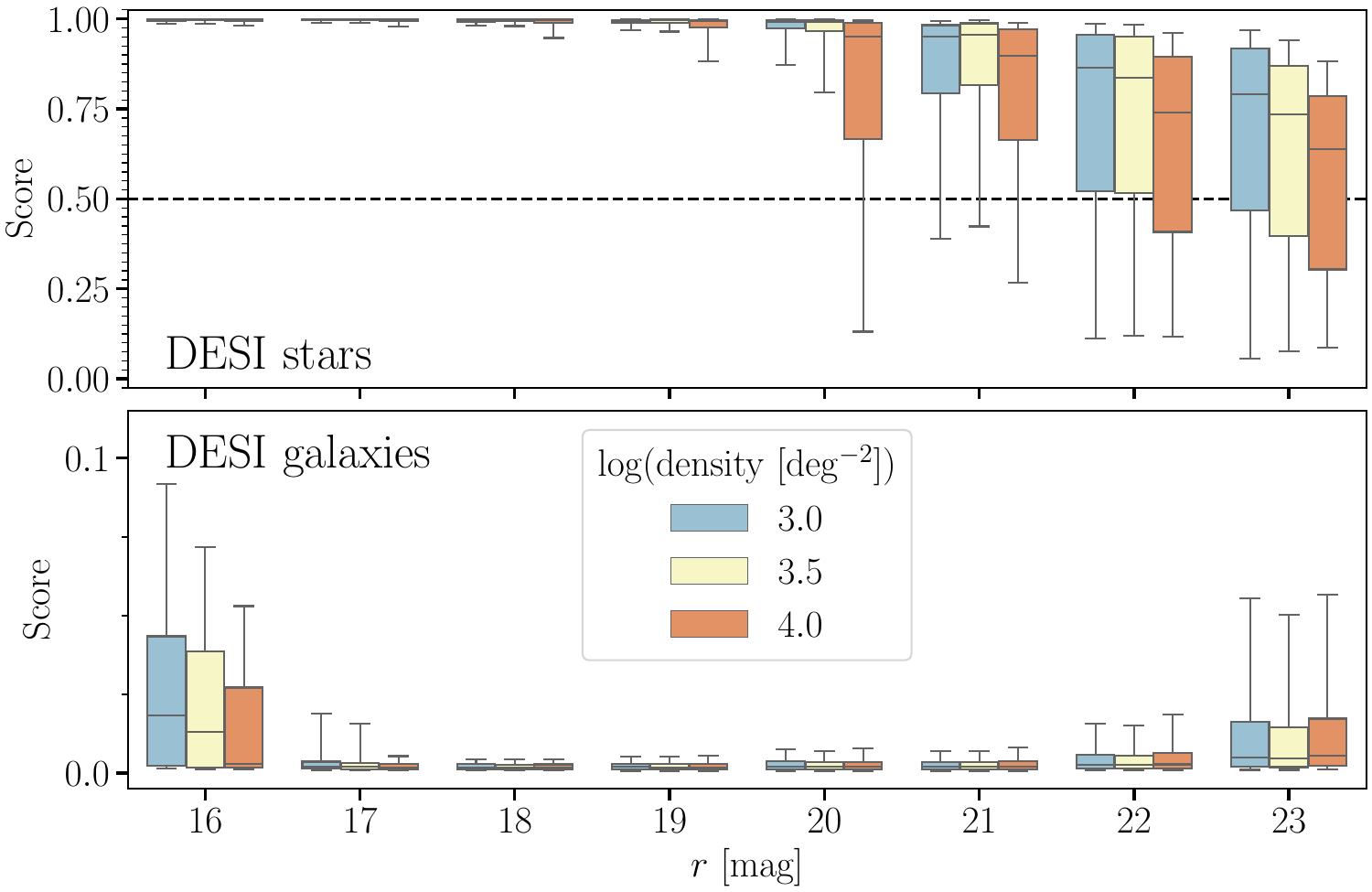}
    \caption{Box plots showing the score distribution for DESI stars and galaxies as a function of stellar crowding. Model scores are lower for stars in fields with higher stellar density, because they are more likely to be blended with another star. The score distribution of galaxies, however, is not affected by the stellar crowding. Stellar density is evaluated using Gaia stars in each LS brick (see Section~\ref{sec:Gaia}), which provides a good proxy for crowding.}
    \label{fig:density}
\end{figure*}
Regions of high stellar density (e.g., the Galactic plane, stellar clusters, local group galaxies) are more likely to produce stellar blends. Blended stars are morphologically similar to extended sources, and as a result our model regularly classifies blended stars as extended sources (rather than 2 or more point sources). As a result, the FNR for our model is sensitive to stellar density, which is not uniform across the sky.

To test model performance as a function of stellar crowding, we show the score distribution for DESI stars and galaxies in fields with different stellar densities in Figure~\ref{fig:density}. The stellar density in each LS brick is evaluated using the number of Gaia stars (defined in Section~\ref{sec:Gaia}).

As stellar crowding increases, the scores of DESI stars decrease. In contrast, the distribution of galaxy scores remains relatively insensitive to crowding, partly because a galaxy blended with a star still appears as an extended source. Consequently, at a fixed classification threshold, we expect a lower TPR in regions with higher stellar density, while the FPR remains largely unaffected. This allows us to provide a single classification threshold with a nearly constant FPR across the sky, with the caveat that the FoM is worse in more crowded fields.

\section{Incorporating the Catalog in LS4}\label{sec:pipeline}
\subsection{LS4 in a Nutshell}
We have presented a classification model to identify resolved and unresolved sources in LS. While it can be applied to more general tasks, our primary goal is to cross match this catalog with the LS4 alert stream, in order to optimize the follow-up strategy of newly discovered transients.

LS4 \citep{LS4_2025} is a new time-domain survey conducted by the ESO 1m Schmidt telescope at La Silla Observatory. It fills the Schmidt focal plane with 32 LBNL deep-depletion well CCDs to provide a total field of view (FoV) of $\sim$20\,$\deg^2$. The LBNL CCDs are especially sensitive to red-optical photons. With the typical 45\,sec exposure, LS4 can reach a depth of $\sim$21.5\,mag in $g$ and $\sim$20.25\,mag in $z$. 
Alerts on newly detected variability in the public surveys will be distributed to the community via Avro data packets in nearly real time.

\subsection{The LS Point Source Catalog}
\begin{figure*}\label{fig:LS-PSC}
    \centering
    \includegraphics[width=0.95\linewidth]{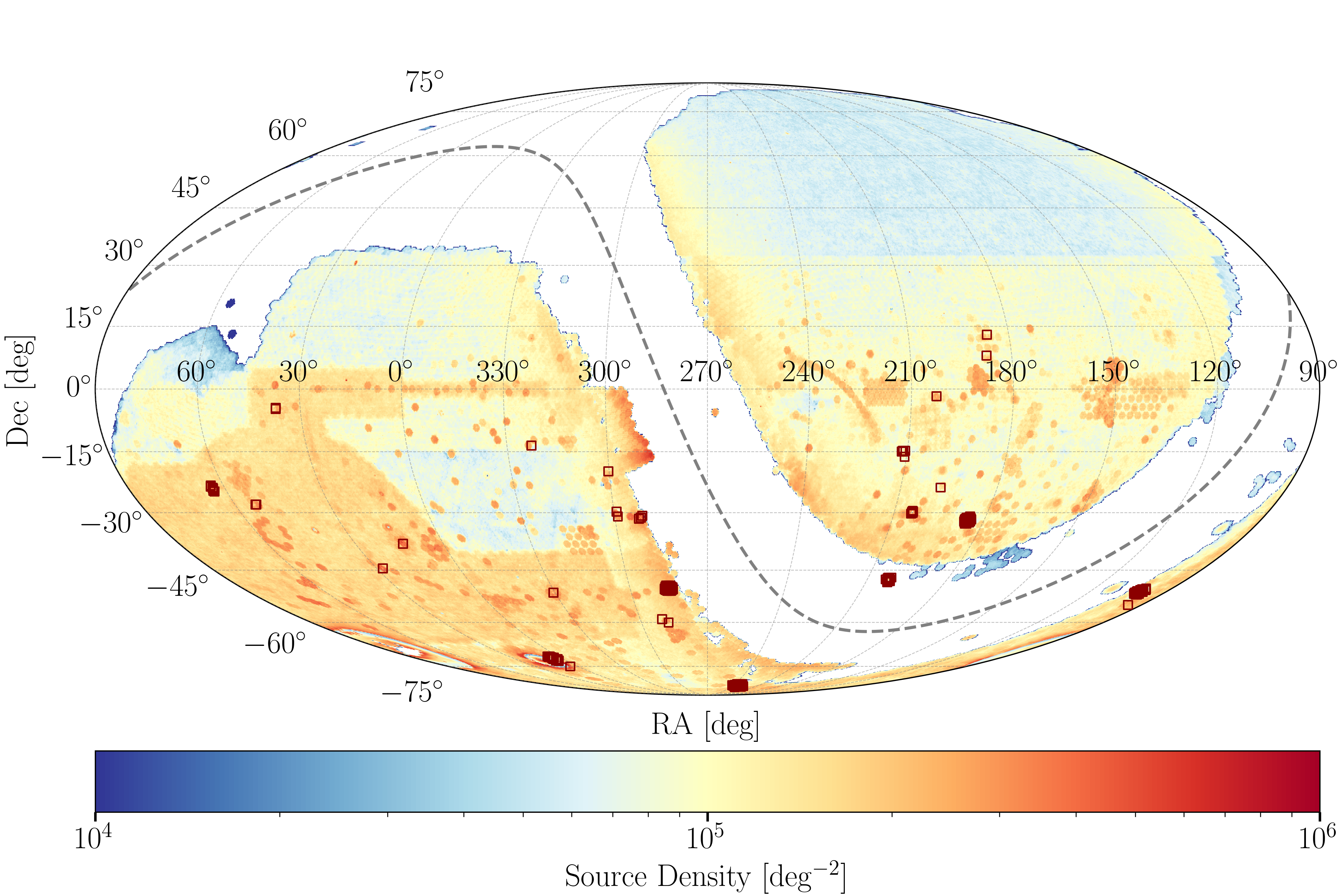}
    \caption{The source density map of LS-PSC. The red boxes indicate the bricks containing sources labeled as \texttt{bailout}. The actual size of the bricks ($\sim$$0.25\degr\times0.25\degr$) is much smaller than the box size. The dash line marks the Galactic plane. The source density in the northern footprint (covered by BASS and MzLS) is significantly lower than that in the south (covered by DECam-based programs such as DECaLS and DES).}
    \label{fig:LS-PSC}
\end{figure*}
Our final data product, the LS Point Source Catalog (LS-PSC), includes 2,807,176,359 LS \dr{10} sources in the south footprint and 339,015,213 sources in the north footprint, for which we provide the scores from the Hybrid model. In Figure~\ref{fig:LS-PSC} we display the all-sky source density distribution. We exclude all DUP sources as well as sources with no valid detections, as illustrated in Section~\ref{sec:data}, when we define the training and test sets. In addition, as a known issue of \dr{10}, source extraction failed on a small portion of bricks (known as \texttt{BAILOUT} regions; see Figure~\ref{fig:LS-PSC} for their locations). In Appendix~\ref{sec:bailout} we describe how we have identified these \texttt{BAILOUT} bricks, from which we are able to retrieve 107,937 \dr{9} sources where there are no \dr{10} sources. In total, there are 3,146,299,509 LS sources with \xgboost\ classifications, making this the largest catalog of resolved--unresolved classifications that has ever been produced. Out of the 3 billion sources there are 9 objects detected in multiple filters, whereas no aperture photometry is provided. By definition (see Equation~\ref{eq:score_hybrid}) they do not have a finite Hybrid score. These sources include (i) 1 bright, saturated galaxy nucleus; (ii) 2 galaxies at the detector edges; and (iii) 6 stars in highly saturated regions (bright stars and/or stellar clusters). We manually assign a score of 0 and 1 to the 3 galaxies and to the 6 stars, respectively.

We recommend 0.5 as the classification threshold, which turns out to be very close to the threshold of a 0.5\% FPR in the HST training set from the COSMOS field ($0.507\pm0.019$) for the Hybrid model we adopt. In Table~\ref{tab:thresh} we present the classification thresholds for a couple of brightness ranges, evaluated on the HST training set with CV, with which users may construct samples of point sources or resolved sources with customized purity and/or completeness. A caveat is that in fields where the stellar density is significantly higher than in the extragalactic COSMOS field (e.g., at lower Galactic latitudes), the TPR will be lower than the values reported in Table~\ref{tab:thresh}. However, as shown in Section~\ref{sec:density}, the FPR remains largely unaffected by stellar crowding.

We further improve the classification (particularly for stars) with Gaia \dr{3}. Every source in the Gaia test set (178,453,652 objects in total; see Section~\ref{sec:Gaia}) is considered as a reliable star, and we add 1 to their ML scores. While \xgboost\ outputs scores between 0 and 1, our catalog includes scores greater than 1. This recovers $\sim$3\% ($\sim$$5\times10^6$) of these Gaia sources which were misclassified as resolved objects by the ML model. Crossing matching with Gaia \dr{3} also effectively eliminates ghosts as residuals in imperfect bright source extraction. Over the entire catalog, 466,195,108 sources are classified as point sources, corresponding to $\sim$15\% of all classifications.

\begin{deluxetable*}{lclccccc}[h]
\tablehead{
    \colhead{Dataset (Size)} & \colhead{Accuracy} & \colhead{FPR} & \colhead{0.005} & \colhead{0.01} & \colhead{0.02} & \colhead{0.05} & \colhead{0.1}
}
\tablecaption{Classification Thresholds for the LS-PSC.\label{tab:thresh}}
\startdata
\multirow{2}{*}{HST Training Set (240,671)} & \multirow{2}{*}{$97.0\pm0.1\%$} & TPR & $0.707\pm0.011$ & $0.748\pm0.007$ & $0.790\pm0.006$ & $0.855\pm0.005$ & $0.904\pm0.005$ \\
&& Threshold & $0.507\pm0.019$ & $0.374\pm0.015$ & $0.248\pm0.011$ & $0.121\pm0.003$ & $0.061\pm0.001$\\
\hline
\multirow{2}{*}{$\texttt{mag\_white}<24$ (134,472)} & \multirow{2}{*}{$97.5\pm0.1\%$} & TPR & $0.842\pm0.008$ & $0.884\pm0.007$ & $0.920\pm0.006$ & $0.953\pm0.006$ & $0.970\pm0.005$\\
&& Threshold & $0.469\pm0.016$ & $0.295\pm0.014$ & $0.154\pm0.004$ & $0.055\pm0.002$ & $0.024\pm0.001$ \\
\hline
\multirow{2}{*}{$\texttt{mag\_white}<22$ (32,999)} & \multirow{2}{*}{$98.9\pm0.1\%$} & TPR & $0.985\pm0.002$ & $0.989\pm0.001$ & $0.992\pm0.002$ & $0.995\pm0.002$ & $0.997\pm0.002$\\
&& Threshold & $0.138\pm0.032$ & $0.064\pm0.003$ & $0.030\pm0.003$ & $0.012\pm0.001$ & $0.006\pm0.001$\\
\enddata
\tablecomments{The values and uncertainties correspond to the means and standard deviations estimated from the 5 folds in CV, where only sources passing the selection criteria (defined in the first column) in each validation set are used for the evaluation.}
\end{deluxetable*}

\subsection{Integration in the Alert Stream}
One of the primary science objectives of LS4 is to discover fast extragalactic transients (e.g., kilonovae) and infant supernovae. The search (and follow-up) for these rapidly evolving transients is inevitably hindered by a large number of foreground, galactic sources (e.g., flaring stars, cataclysmic variables). A resolved--unresolved model will identify stars/galaxies associated with LS4 transients as solid evidence of their galactic/extragalactic origins. LS-PSC provides a list of galactic sources to eliminate in extragalactic surveys, while keeping almost all galaxies ($\sim$99.5\% when adopting the recommended threshold of 0.5). Specially, our model is much deeper than typical LS4 observations ($\lesssim$21\,mag) and can effectively uncover faint stars up to $\sim$23\,mag that will not be detectable even in stacked reference images for LS4.

Similar to the ZTF alert stream \citep{ZTF_data_2019}, we associate newly discovered LS4 transients with a small number of the nearest LS counterparts within 30\arcsec. For every counterpart found in the cross match, the angular separation from the transient, catalog ID (\texttt{ls\_id}), brightness (\texttt{white\_mag}), and the point-source score will be presented in the real-time alert packets.

\section{Summary and Conclusions}\label{sec:conclusion}
We have presented our morphological model to separate resolved--unresolved sources in LS \dr{10} imaging providing ML scores for $\sim$$3\times10^9$ sources ($\lesssim$26\,mag). The classifier is built based on the Gradient Boosting Decision Tree framework, \xgboost, and is trained using $\sim$$2\times10^6$ LS sources in the COSMOS field with HST morphological labels \citep{Leauthaud_2007}. The model features are derived from the \texttt{Tractor} photometry and LS aperture photometry products. 

Generally speaking, the \xgboost\ models show a higher TPR at constant FPR than the LS morphological model. At a classification threshold of 0.5, our \xgboost\ models successfully recover almost all galaxies but miss the majority of faint stars ($\gtrsim$23\,mag), whereas the LS model is more sensitive to stars but misidentifies a substantial fraction of galaxies. Since the number of observed stars and galaxies are intrinsically highly imbalanced (i.e., dominated by faint galaxies), the \xgboost\ models generally achieve significantly higher classification accuracy than the LS morphological typing. The high spatial resolution and near-infrared capabilities of future space-based surveys (e.g., Roman) will be crucial for distinguishing the faintest stars -- primarily M-dwarfs -- from background galaxies.

We evaluate the model performance on LS sources with spectroscopic classifications from DESI \dr{1} ($\sim$$10^7$ objects) and SDSS ($\sim$$3\times10^6$ objects). The inclusion of aperture flux ratios from individual filters ($grz$) as model features can boost the classification accuracy when data from all filters are available, and observations from the redder filters are generally of higher importance. However, features containing the ``white'' aperture flux, i.e., by combining aperture flux measured in all available filters, improve the model's robustness against missing values. To leverage the strengths of these different approaches, we construct a Hybrid model as a linear combination of two \xgboost\ models, each containing features combining aperture flux measurements from the ``blue'' ($gr$) and ``red'' ($iz$) filters. On the HST training set, it achieves an FoM of $0.707\pm0.011$ in CV, outperforming the White model ($0.668\pm0.012$) containing only the ``white'' aperture flux ratios. On the DESI and SDSS test sets it outperforms both the White and $grz$ models in FoM. Ultimately, the Hybrid model achieves the best balance between sensitivity and robustness, and is adopted in building our final point-source catalog, LS-PSC.

LS-PSC is integrated into the real-time transient detection pipeline for LS4, which will produce public alerts in mid-2025. Our catalog enables the systematic identification and removal of foreground Galactic transients and variables, a major source of false positives in the search for extragalactic transients. The catalog is available at \url{https://ls-xgboost.lbl.gov} via API requests.

\software{
    \texttt{Astropy} \citep{Astropy_2013,Astropy_2018},
    \texttt{matplotlib} \citep{Matplotlib_2007}, 
    \texttt{NumPy} \citep{numpy_2020},
    \texttt{Pandas} \citep{Pandas_2010},
    \texttt{scikit-learn} \citep{scikit-learn},
    \texttt{seaborn} \citep{Waskom_seaborn_2021},
    \xgboost\ \citep{XGBoost_2016}
}

\facility{Astro Data Lab, Gaia}

\section*{Acknowledgement}
{We thank the anonymous referee for the constructive comments to improve the paper.} We thank Chris Stubbs for fruitful discussions. We thank Robert Nikutta and Stephanie Juneau for the aid of accessing the LS \dr{10} catalog, and Dustin Lang for explaining some details in the LS data processing pipeline. We thank Micheal Stroh for the support in the Northwestern high-performance computing cluster Quest. C.L. and~A.A.M~are supported by DoE award \#DE-SC0025599.

This research was supported in part through the computational resources and staff contributions provided for the Quest high performance computing facility at Northwestern University which is jointly supported by the Office of the Provost, the Office for Research, and Northwestern University Information Technology. 

This research used resources of the National Energy Research Scientific Computing Center (NERSC), a Department of Energy Office of Science User Facility using NERSC award HEP-ERCAP-0033561. P.E.N. acknowledges support from the DOE/ASCR
through DE-FOA-0001088, Analytical Modeling for Extreme-Scale Computing Environments, the X-SWAP Project. 

This research uses services or data provided by the Astro Data Lab, which is part of the Community Science and Data Center (CSDC) Program of NSF NOIRLab.

The Legacy Surveys consist of three individual and complementary projects: the Dark Energy Camera Legacy Survey (DECaLS; Proposal ID \#2014B-0404; PIs: David Schlegel and Arjun Dey), the Beijing-Arizona Sky Survey (BASS; NOAO Prop. ID \#2015A-0801; PIs: Zhou Xu and Xiaohui Fan), and the Mayall z-band Legacy Survey (MzLS; Prop. ID \#2016A-0453; PI: Arjun Dey). DECaLS, BASS and MzLS together include data obtained, respectively, at the Blanco telescope, Cerro Tololo Inter-American Observatory, NSF's NOIRLab; the Bok telescope, Steward Observatory, University of Arizona; and the Mayall telescope, Kitt Peak National Observatory, NOIRLab. Pipeline processing and analyses of the data were supported by NOIRLab and the Lawrence Berkeley National Laboratory (LBNL). The Legacy Surveys project is honored to be permitted to conduct astronomical research on Iolkam Du'ag (Kitt Peak), a mountain with particular significance to the Tohono O'odham Nation.

NOIRLab is operated by the Association of Universities for Research in Astronomy (AURA) under a cooperative agreement with the National Science Foundation. LBNL is managed by the Regents of the University of California under contract to the U.S. Department of Energy.

This project used data obtained with the Dark Energy Camera (DECam), which was constructed by the Dark Energy Survey (DES) collaboration. Funding for the DES Projects has been provided by the U.S. Department of Energy, the U.S. National Science Foundation, the Ministry of Science and Education of Spain, the Science and Technology Facilities Council of the United Kingdom, the Higher Education Funding Council for England, the National Center for Supercomputing Applications at the University of Illinois at Urbana-Champaign, the Kavli Institute of Cosmological Physics at the University of Chicago, Center for Cosmology and Astro-Particle Physics at the Ohio State University, the Mitchell Institute for Fundamental Physics and Astronomy at Texas A\&M University, Financiadora de Estudos e Projetos, Fundacao Carlos Chagas Filho de Amparo, Financiadora de Estudos e Projetos, Fundacao Carlos Chagas Filho de Amparo a Pesquisa do Estado do Rio de Janeiro, Conselho Nacional de Desenvolvimento Cientifico e Tecnologico and the Ministerio da Ciencia, Tecnologia e Inovacao, the Deutsche Forschungsgemeinschaft and the Collaborating Institutions in the Dark Energy Survey. The Collaborating Institutions are Argonne National Laboratory, the University of California at Santa Cruz, the University of Cambridge, Centro de Investigaciones Energeticas, Medioambientales y Tecnologicas-Madrid, the University of Chicago, University College London, the DES-Brazil Consortium, the University of Edinburgh, the Eidgenossische Technische Hochschule (ETH) Zurich, Fermi National Accelerator Laboratory, the University of Illinois at Urbana-Champaign, the Institut de Ciencies de l'Espai (IEEC/CSIC), the Institut de Fisica d'Altes Energies, Lawrence Berkeley National Laboratory, the Ludwig Maximilians Universitat Munchen and the associated Excellence Cluster Universe, the University of Michigan, NSF's NOIRLab, the University of Nottingham, the Ohio State University, the University of Pennsylvania, the University of Portsmouth, SLAC National Accelerator Laboratory, Stanford University, the University of Sussex, and Texas A\&M University.

BASS is a key project of the Telescope Access Program (TAP), which has been funded by the National Astronomical Observatories of China, the Chinese Academy of Sciences (the Strategic Priority Research Program “The Emergence of Cosmological Structures” Grant \# XDB09000000), and the Special Fund for Astronomy from the Ministry of Finance. The BASS is also supported by the External Cooperation Program of Chinese Academy of Sciences (Grant \# 114A11KYSB20160057), and Chinese National Natural Science Foundation (Grant \# 12120101003, \# 11433005).

The Legacy Survey team makes use of data products from the Near-Earth Object Wide-field Infrared Survey Explorer (NEOWISE), which is a project of the Jet Propulsion Laboratory/California Institute of Technology. NEOWISE is funded by the National Aeronautics and Space Administration.

The Legacy Surveys imaging of the DESI footprint is supported by the Director, Office of Science, Office of High Energy Physics of the U.S. Department of Energy under Contract No. DE-AC02-05CH1123, by the National Energy Research Scientific Computing Center, a DOE Office of Science User Facility under the same contract; and by the U.S. National Science Foundation, Division of Astronomical Sciences under Contract No. AST-0950945 to NOAO.




\bibliography{main,software,facility}{}
\bibliographystyle{aasjournal}

\appendix

\section{Feature Importance}\label{sec:feature_importance}

In this section we report the feature importance for each \xgboost\ model, defined as the average gain in accuracy when a feature is used in a split. The results are displayed in Table~\ref{tab:feature_importance_white} (White model), Table~\ref{tab:feature_importance_grz} ($grz$ model), and Table~\ref{tab:feature_importance_hyb} (Hybrid model). Features are sorted by descending importance; those with equal importance are then sorted alphabetically. For the Hybrid model, we report the feature importance for both \xgboost\ models.

{
\begin{deluxetable}{lc}[h]\label{tab:feature_importance_white}
\tabletypesize{\footnotesize}
\tablehead{
    \colhead{Name} & \colhead{White}
}
\tablecaption{The feature importance for the White model.}
\startdata
\texttt{mask} & $0.38 \pm 0.07$ \\
\texttt{delta\_dchisq}(REX) & $0.218 \pm 0.025$ \\
\texttt{delta\_dchisq}(EXP) & $0.118 \pm 0.013$ \\
\texttt{delta\_dchisq}(DEV) & $0.056 \pm 0.006$ \\
\texttt{apratio\_white}(2\farcs0, 3\farcs5) & $0.056 \pm 0.008$ \\
\texttt{apratio\_white}(1\farcs5, 2\farcs0) & $0.033 \pm 0.003$ \\
\texttt{apratio\_white}(5\farcs0, 7\farcs0) & $0.027 \pm 0.004$ \\
\texttt{apratio\_white}(3\farcs5, 5\farcs0) & $0.026 \pm 0.002$ \\
\texttt{delta\_dchisq}(SER) & $0.022 \pm 0.002$ \\
\texttt{apratio\_white}(0\farcs5, 0\farcs75) & $0.020 \pm 0.002$ \\
\texttt{apratio\_white}(1\farcs0, 1\farcs5) & $0.020 \pm 0.002$ \\
\texttt{apratio\_white}(0\farcs75, 1\farcs0) & $0.019 \pm 0.002$ \\
\enddata
\end{deluxetable}

\begin{deluxetable}{lc}[h]\label{tab:feature_importance_grz}
\tabletypesize{\footnotesize}
\tablehead{
    \colhead{Name} & \colhead{$grz$}
}
\tablecaption{The feature importance for the $grz$ model.}
\startdata
\texttt{mask} & $0.54 \pm 0.03$ \\
\texttt{delta\_dchisq}(EXP) & $0.099 \pm 0.006$ \\
\texttt{delta\_dchisq}(REX) & $0.065 \pm 0.004$ \\
\texttt{delta\_dchisq}(DEV) & $0.039 \pm 0.003$ \\
\texttt{apratio\_z}(2\farcs0, 3\farcs5) & $0.0320 \pm 0.0025$ \\
\texttt{delta\_dchisq}(SER) & $0.021 \pm 0.002$ \\
\texttt{apratio\_z}(1\farcs0, 1\farcs5) & $0.0188 \pm 0.0014$ \\
\texttt{apratio\_r}(0\farcs5, 0\farcs75) & $0.0185 \pm 0.0024$ \\
\texttt{apratio\_z}(1\farcs5, 2\farcs0) & $0.0177 \pm 0.0012$ \\
\texttt{apratio\_z}(5\farcs0, 7\farcs0) & $0.0134 \pm 0.0008$ \\
\texttt{apratio\_r}(0\farcs75, 1\farcs0) & $0.0126 \pm 0.0021$ \\
\texttt{apratio\_r}(2\farcs0, 3\farcs5) & $0.0124 \pm 0.0016$ \\
\texttt{apratio\_z}(0\farcs75, 1\farcs0) & $0.0122 \pm 0.0007$ \\
\texttt{apratio\_g}(0\farcs5, 0\farcs75) & $0.0109 \pm 0.0007$ \\
\texttt{apratio\_z}(3\farcs5, 5\farcs0) & $0.0106 \pm 0.0009$ \\
\texttt{apratio\_g}(1\farcs0, 1\farcs5) & $0.0099 \pm 0.0008$ \\
\texttt{apratio\_z}(0\farcs5, 0\farcs75) & $0.0093 \pm 0.0007$ \\
\texttt{apratio\_g}(0\farcs75, 1\farcs0) & $0.0088 \pm 0.0006$ \\
\texttt{apratio\_g}(1\farcs5, 2\farcs0) & $0.0078 \pm 0.0005$ \\
\texttt{apratio\_r}(3\farcs5, 5\farcs0) & $0.0074 \pm 0.0007$ \\
\texttt{apratio\_r}(1\farcs5, 2\farcs0) & $0.0073 \pm 0.0005$ \\
\texttt{apratio\_g}(2\farcs0, 3\farcs5) & $0.0072 \pm 0.0004$ \\
\texttt{apratio\_r}(1\farcs0, 1\farcs5) & $0.0072 \pm 0.0005$ \\
\texttt{apratio\_r}(5\farcs0, 7\farcs0) & $0.0064 \pm 0.0005$ \\
\texttt{apratio\_g}(3\farcs5, 5\farcs0) & $0.0063 \pm 0.0004$ \\
\texttt{apratio\_g}(5\farcs0, 7\farcs0) & $0.0062 \pm 0.0005$ \\
\enddata
\end{deluxetable}

\begin{deluxetable}{lc|lc}\label{tab:feature_importance_hyb}
\tabletypesize{\footnotesize}
\tablehead{
    \colhead{Name} & \colhead{Hybrid ($gr$)} &\colhead{Name} & \colhead{Hybrid ($iz$)}
}
\tablecaption{The feature importance for the Hybrid model.}
\startdata
\texttt{mask} & $0.52 \pm 0.07$ & \texttt{delta\_dchisq}(REX) & $0.315 \pm 0.034$ \\
\texttt{delta\_dchisq}(REX) & $0.192 \pm 0.029$ & \texttt{mask} & $0.17 \pm 0.09$ \\
\texttt{delta\_dchisq}(EXP) & $0.082 \pm 0.013$ & \texttt{delta\_dchisq}(EXP) & $0.136 \pm 0.013$ \\
\texttt{delta\_dchisq}(DEV) & $0.050 \pm 0.007$ & \texttt{apratio\_iz}(2\farcs0, 3\farcs5) & $0.076 \pm 0.008$ \\
\texttt{apratio\_gr}(2\farcs0, 3\farcs5) & $0.024 \pm 0.003$ & \texttt{delta\_dchisq}(DEV) & $0.054 \pm 0.008$ \\
\texttt{delta\_dchisq}(SER) & $0.019 \pm 0.003$ & \texttt{apratio\_iz}(1\farcs5, 2\farcs0) & $0.050 \pm 0.007$ \\
\texttt{apratio\_gr}(5\farcs0, 7\farcs0) & $0.019 \pm 0.003$ & \texttt{apratio\_iz}(5\farcs0, 7\farcs0) & $0.041 \pm 0.004$ \\
\texttt{apratio\_gr}(0\farcs5, 0\farcs75) & $0.018 \pm 0.003$ & \texttt{apratio\_iz}(3\farcs5, 5\farcs0) & $0.036 \pm 0.003$ \\
\texttt{apratio\_gr}(0\farcs75, 1\farcs0) & $0.018 \pm 0.003$ & \texttt{apratio\_iz}(0\farcs5, 0\farcs75) & $0.035 \pm 0.003$ \\
\texttt{apratio\_gr}(1\farcs5, 2\farcs0) & $0.018 \pm 0.003$ &  \texttt{apratio\_iz}(1\farcs0, 1\farcs5) & $0.035 \pm 0.003$\\
\texttt{apratio\_gr}(3\farcs5, 5\farcs0) & $0.018 \pm 0.003$ & \texttt{apratio\_iz}(0\farcs75, 1\farcs0) & $0.031 \pm 0.003$ \\
\texttt{apratio\_gr}(1\farcs0, 1\farcs5) & $0.017 \pm 0.003$ & \texttt{delta\_dchisq}(SER) & $0.028 \pm 0.003$ \\
\enddata
\end{deluxetable}
}

\section{Handling \texttt{BAILOUT} Bricks in LS \dr{10}}\label{sec:bailout}
One known issue of LS \dr{10} is that during source extraction, some bricks, typically those with exceptionally high source density, cannot be fully processed in a reasonable timescale. The pipeline ``bailed out'' of these bricks before they finished processing. As a result, many objects from these bricks were not included in the catalog. Sources successfully extracted from these regions have their \texttt{BAILOUT} flag set in the \texttt{MASKBITS} bitmask, with which we can identify the bailout fields. In total, we find 206 bricks with at least one source flagged as \texttt{BAILOUT}, totaling $\sim$$12.8\,\deg^2$, or $\sim$0.056\% of the \dr{10} footprint, although typically only part of the brick is bailed out. Therefore up to $\sim$$10^6$ sources might be missing in the LS \dr{10} catalog.

Some of these \texttt{BAILOUT} bricks overlap the \dr{9} footprint, where the sources should be cataloged in \dr{9} (though extracted from shallower images). To prevent large ``holes'' in the sky coverage of our catalog, we retrieve the \dr{9} sources with no corresponding \dr{10} sources in these \texttt{BAILOUT} bricks. 

To ensure completeness while searching for \texttt{BAILOUT} fields in \dr{10}, we include both the 206 bricks with \texttt{BAILOUT} sources, as well as their adjacent bricks (i.e., the row number \texttt{brickrow} and column number \texttt{brickcol} differ by no more than 1). This adds up to 803 candidate \texttt{BAILOUT} bricks, where there are 143 bricks in the \dr{9} footprint. We then query all sources from the 143 bricks, yielding 1,093,455 \dr{9} sources and 2,193,130 \dr{10} sources. Among all the \dr{9} sources, 119,019 do not have a counterpart in \dr{10} within 3\farcs0. A small number ($\sim$$10$ per brick) of them are probably artifacts due to problematic photometry of \dr{9} imaging, whereas the vast majority are real sources in the fields that were bailed out of in \dr{10}. By visually inspecting all bricks with more than 100 unmatched \dr{9} sources, we identify 25 bricks (\texttt{brickid} = 60861,  71239,  72134,  73616,  73940,  98957, 127083, 174243, 175512, 175513, 175514, 191029, 192338, 193649, 194961, 196277, 197595, 198917, 200239, 254534, 303174, 304610, 321098, 377081,
405415) with significant continuous regions where no sources are extracted, including 3 bricks (\texttt{brickid} = 174243, 175514, 191029) that are included as adjacent bricks of the original bricks, meaning they happen to not contain any \dr{10} sources flagged as \texttt{BAILOUT}. There are 107,937 unmatched \dr{9} sources in those 25 bricks, which we include as part of our final catalog. 

\end{CJK*}
\end{document}